\documentclass[nofootinbib, aps, prd, a4paper, 10pt, superscriptaddress, eqsecnum, showkeys]{revtex4-2}
\usepackage[a4paper, top=2cm, bottom=2cm, left=2.5cm, right=2.5cm]{geometry}

\usepackage{amsmath}
\usepackage{amsfonts}
\usepackage{amsthm}
\usepackage{bm}
\usepackage{mathrsfs}
\usepackage{braket}

\usepackage{graphicx}
\usepackage{booktabs}

\usepackage{xcolor}
\usepackage[pdfusetitle]{hyperref}
\hypersetup{colorlinks = true, allcolors = blue}

\usepackage{orcidlink}

\begin{document}

\title{String-inspired Gauss-Bonnet Gravity Inflation and ACT}

\author{S.D.~Odintsov\orcidlink{0000-0002-3529-7030}}
\email{odintsov@ice.cat} \affiliation{ICREA, Passeig Luis
Companys, 23, 08010 Barcelona, Spain} \affiliation{Institute of
Space Sciences (IEEC-CSIC) C. Can Magrans s/n, 08193 Barcelona,
Spain} \affiliation{L.N. Gumilyov Eurasian National University -
Astana, 010008, Kazakhstan}

\author{V.K.~Oikonomou\orcidlink{0000-0003-0125-4160}}
\email{voikonomou@gapps.auth.gr} \affiliation{Department of
Physics, Aristotle University of Thessaloniki, Thessaloniki 54124,
Greece} \affiliation{Center for Theoretical Physics, Khazar
University, 41 Mehseti Str., Baku, AZ-1096, Azerbaijan}
\affiliation{L.N. Gumilyov Eurasian National University - Astana,
010008, Kazakhstan}

\author{Pyotr~Tsyba\orcidlink{0000-0003-4928-0392}}
\email{pyotrtsyba@gmail.com} \affiliation{L.N. Gumilyov Eurasian
National University - Astana, 010008, Kazakhstan}

\author{Olga~Razina\orcidlink{0000-0002-4400-4789}}
\email{olvikraz@mail.ru} \affiliation{L.N. Gumilyov Eurasian
National University - Astana, 010008, Kazakhstan}

\author{Dauren Rakhatov}\orcidlink{0000-0002-5318-2990}
\email{godauren@gmail.com} \affiliation{L.N. Gumilyov Eurasian
National University - Astana, 010008, Kazakhstan}


\begin{abstract}
In this article we present a systematic observational verification
of the ghost-free string-inspired $f(R,\mathcal{G})$ model, where
the Gauss-Bonnet invariant is non-minimally coupled to an
auxiliary scalar field $\chi$ through the coupling function
$h(\chi)$. Previous studies confirmed the theoretical viability of
this framework using phenomenological parameter choices. In this
work, for the first time, a systematic comparison with
observational data from Planck 2018 and the Atacama Comsology
Telescope is carried out via a Bayesian MCMC analysis using the
Cobaya code. We explore an extended set of sixteen models
constructed from four types of the Hubble parameter combined with
power-law, exponential, hybrid, and inverse logarithmic coupling
functions $h(\chi)$. The hybrid coupling $h(\chi) = \gamma
e^{b_1\chi}\chi^{b_2}$, introduced in this context, allows for
interpolation between the power-law and exponential forms,
providing additional flexibility in controlling the Gauss-Bonnet
contribution at different stages of inflation. All sixteen models
reproduce the red spectral tilt of scalar perturbations consistent
with CMB observations, yielding $n_s \approx 0.97$ at $N = 60$
e-folds. We find that the preference for the dataset is
systematically determined by the choice of Hubble parametrization
rather than by the coupling function. The parameter
$\mu\approx0.1$ remains stable in all configurations, suggesting
its fundamental role within the ghost-free formalism.
\end{abstract}

\maketitle

\section{Introduction}

The scenario of early inflationary expansion was proposed to solve
the problems of horizon, flatness, and magnetic monopoles
\cite{inflation1,inflation2,inflation3,inflation4}. According to
this scenario, the early Universe exponentially increased in size
by many orders of magnitude over a negligibly short period of
time. Unfortunately, getting direct data from that era is
problematic. But despite this, inflation generates a primordial
spectrum of scalar and tensor perturbations, observable as
anisotropies in the cosmic microwave background (CMB) power
spectrum. The key observational predictions of inflation are the
red tilt of the scalar perturbation spectrum ($n_s < 1$) and a
nonzero tensor-to-scalar ratio $r$, the measurement of which
allows competing inflationary models to be distinguished.

The most common mechanism for implementing the inflationary
scenario is the introduction of a scalar field. However, scalar
fields are not the only method of obtaining a viable inflationary
stage. Alternative approaches are being actively developed and
implemented through modified theories of gravity
\cite{reviews1,reviews2,reviews3,reviews4}, which represent a rich
class of inflationary models. Among them, a special place is
occupied by the family of $f(R)$ theories of gravity, in which the
standard Einstein-Hilbert action is modified by an arbitrary
function of the Ricci scalar. Further generalizations lead to the
inclusion of the Gauss-Bonnet invariant $\mathcal{G}$ with
quadratic curvature terms, which naturally arises in the
low-energy limit of string theory
\cite{Hwang:2005hb,Nojiri:2006je,Cognola:2006sp,Nojiri:2005vv,Nojiri:2005jg,Satoh:2007gn,
Bamba:2014zoa,Yi:2018gse,Guo:2009uk,Guo:2010jr,Jiang:2013gza,vandeBruck:2017voa,Pozdeeva:2020apf,
Vernov:2021hxo,Pozdeeva:2021iwc,Fomin:2020hfh,DeLaurentis:2015fea,Chervon:2019sey,Nozari:2017rta,
Odintsov:2018zhw,Kawai:1998ab,Yi:2018dhl,vandeBruck:2016xvt,Maeda:2011zn,Ai:2020peo,Easther:1996yd,Codello:2015mba,
Oikonomou:2021kql,Odintsov:2020sqy,Oikonomou:2024etl,Fier:2025huc,Oikonomou:2022ksx,Oikonomou:2022xoq}.
Combined theories of $f(R,\mathcal{G})$ gravity provide a rich
phenomenology, showing the possibility of describing the early and
late stages of the Universe's evolution.

However, higher-derivative theories, including $f(\mathcal{G})$
and $f(R,\mathcal{G})$ models, may suffer from instabilities. The
presence of second-order derivatives in Lagrangians leads to
equations of motion containing fourth-order derivatives, which
results in degrees of freedom with negative kinetic energy.
Instabilities of this kind are called Ostrogradsky's ghosts, the
presence of which violates the unitarity of theory and is a
serious theoretical obstacle to building viable models. A solution
to this problem was proposed in
\cite{Nojiri:2021mxf,Nojiri:2019dwl}, where a mechanism was
developed to suppress ghost degrees of freedom by introducing a
Lagrange multiplier $\lambda$ and an auxiliary scalar field $\chi$
with a mass dimension. Viable inflationary scenarios were
investigated in \cite{Nojiri:2021mxf,Nojiri:2019dwl}, where the
possibility of describing the early Universe was demonstrated on
the basis of three types of cosmological evolution. In this model,
the contribution of the Gauss-Bonnet invariant is controlled by a
non-minimal coupling with the auxiliary scalar field $h(\chi)$.
The use of various coupling functions makes it possible to
implement a wide range of consistent inflationary scenarios of the
model.

Over the past decade, many inflationary models have been
thoroughly tested, limited by the precession CMB measurements of
the Planck collaboration \cite{Planck:2018jri}. According to the
latest Planck 2018 release, the spectral index of scalar
perturbations is $n_s=0.9649\pm 0.0042$, and the tensor-to-scalar
ratio is bounded from above by $r < 0.036$ together with
BICEP/Keck data \cite{BICEP:2021xfz}. These results have become a
benchmark for testing inflationary scenarios and have led to the
rejection of a significant number of models. The recent sixth
release (DR6.02) of data from the Atacama Cosmology Telescope
(ACT) has opened up new research opportunities by providing
high-resolution measurements of polarization spectra at small
angular scales \cite{ACT:2025fju, ACT:2025tim}. A joint analysis
of the Planck and ACT data revealed a partial discrepancy in the
$(n_s,r)$ plane, with ACT preferring slightly higher values of the
spectral index of scalar perturbations. This raises the question
of which inflationary models are able to satisfy both datasets,
and what the preferences of the different models are between the
two datasets. After the ACT data release, many articles appeared
in the literature, trying to reconcile standard inflation
scenarios with the ACT findings
\cite{Kallosh:2025rni,Gao:2025onc,Liu:2025qca,Yogesh:2025wak,Yi:2025dms,Peng:2025bws,Yin:2025rrs,Byrnes:2025kit,
Wolf:2025ecy,Aoki:2025wld,Gao:2025viy,Zahoor:2025nuq,Ferreira:2025lrd,Mohammadi:2025gbu,Choudhury:2025vso,
Odintsov:2025wai,Q:2025ycf,Zhu:2025twm,Kouniatalis:2025orn,Hai:2025wvs,Dioguardi:2025vci,Yuennan:2025kde,
Kuralkar:2025zxr,Kuralkar:2025hoz,Modak:2025bjv,Oikonomou:2025xms,Odintsov:2025jky,
Aoki:2025ywt,Ahghari:2025hfy,McDonough:2025lzo,Chakraborty:2025wqn,NooriGashti:2025gug,Yuennan:2025mlg,
Deb:2025gtk,Afshar:2025ndm,Ellis:2025zrf,Yuennan:2025tyx,Wang:2025cpp,Qiu:2025uot,Wang:2025dbj,Asaka:2015vza,Oikonomou:2025htz,Choudhury:2025hnu,Singh:2025uyr,Kim:2025dyi,Peng:2026ofs,Yuennan:2026fcn,Yogesh:2026esn}

Despite the theoretical validity of ghost-free $f(R,\mathcal{G})$
gravity, its inflationary scenarios have been studied within the
framework of a reconstructive approach with a phenomenological
selection of model parameters. This approach makes it possible to
effectively demonstrate the feasibility of various cosmological
scenarios, but does not imply a strict statistical verification
based on real observational data. The present work develops this
approach, complementing it with a systematic Bayesian MCMC
analysis based on Planck and ACT data, moving to dimensionless
normalization $H_0=1$ and natural units $\kappa^2=1$, ensuring
consistency of parameters with physically motivated scales. The
study examines sixteen models based on four types of evolution of
the Hubble parameter together with four types of the coupling
function $h(\chi)$. For the first time, the hybrid potential
$h(\chi) = \gamma\mathrm{e}^{b_1\chi}\chi^{b_2}$ was proposed and
investigated, allowing balancing between the power-law and
exponential cases, which provides flexibility in adjusting the
effect of the Gauss-Bonnet invariant at various stages of
inflation. A central objective of this work is the verification
and systematic comparison of the preferences of each model between
the Planck and ACT data.

The article is organized as follows. Section~\ref{sec:theory}
outlines the theoretical basis of the model, including action,
equations of motion, and inflationary observables.
Section~\ref{sec:method} describes the methodology of Bayesian
analysis. The results for the sixteen models are presented in
sections~\ref{sec:dS}-\ref{sec:frac}. The summary and conclusions
are contained in the section~\ref{sec:discussion}.

\section{Theoretical Framework}\label{sec:theory}

\subsection{Action and Equations of Motion}

The ghost-free $f(R,\mathcal{G})$ gravity is described by the
action \cite{Nojiri:2021mxf,Nojiri:2019dwl},
\begin{equation}\label{eq:action}
    S = \int d^4x \sqrt{-g} \left[ \frac{R}{2\kappa^2} + \frac{\lambda}{2} \left( \partial_\mu \chi \partial^\mu \chi + \mu^4 \right) - \frac{1}{2} \partial_\mu \chi \partial^\mu \chi + h(\chi) \mathcal{G} - V(\chi) + L_m  \right],
\end{equation}
where $g$ is the determinant of the metric tensor $g_{\mu\nu}$,
$R$ is the Ricci scalar, $\kappa^2 = 8\pi G$ is the gravitational
coupling constant, $\lambda$ is the Lagrange multiplier enforcing
the ghost-free constraint, $\chi$ is the auxiliary scalar field
with mass-dimension constant $\mu$, $h(\chi)$ is the coupling
function between the scalar field and the Gauss-Bonnet invariant
$\mathcal{G} = R_{\mu\nu\rho\sigma}R^{\mu\nu\rho\sigma} -
4R_{\mu\nu}R^{\mu\nu} + R^2$, $V(\chi)$ is the scalar field
potential, and $L_m$ is the matter Lagrangian. This action without
second Lagrange multiplier-term is a typical string-inspired
gravity which results from bosonic or Type II string up to third
order term. However, it is well-known that such third-order string
inspired action may lead to ghosts and other instabilities. That
is reason to introduce the Lagrange-multiplier term which makes
the theory to be ghost free. That is the reason why Lagrange
multiplier term is added.

Varying the action (\ref{eq:action}) with respect to the Lagrange
multiplier $\lambda$, results in the constraint equation
$\partial_\mu \chi \partial^\mu \chi + \mu^4 = 0$, leading to a
constant kinetic term, which allows for defining the effective
potential,
\begin{equation}
    \tilde{V}(\chi) = \frac{1}{2}\partial_\mu \chi \partial^\mu \chi + V(\chi) = -\frac{\mu^4}{2} + V(\chi).
\end{equation}

In the flat Friedmann-Robertson-Walker background,
$ds^2=-dt^2+a(t)^2(dx^2+dy^2+dz^2)$, variation of the action
(\ref{eq:action}) with respect to the metric and the scalar field
gives the equations of motion,
\begin{align}
\label{eq:F1norm}
&\frac{1}{2\kappa^2} \big(3H^2 + 2\dot{H}\big)
+ 8\mu^2 H \big( H^2+\dot{H} \big)h'(\chi)
+ 4\mu^4 H^2 h''(\chi)
- \frac{1}{2}\tilde{V}(\chi)
= 0, \\
\label{eq:F2norm}
&-\frac{3H^2}{2\kappa^2}
+ \frac{1}{2} \tilde{V} (\chi)
- \frac{1}{4} \mu^4 \lambda
- 12 \mu^2 H^3 h'(\chi)
= 0, \\
\label{eq:KGnorm}
&\mu^2 \dot{\lambda}
+ 3\mu^2 H\lambda
+ \mathcal{G}h'(\chi)
- \tilde{V}'(\chi)
= 0.
\end{align}
where $H$ is the Hubble parameter, the ``dots'' denote derivatives
with respect to cosmic time $t$, and the ``primes'' denote
derivatives with respect to the auxiliary field $\chi = \mu^2 t$.
Combining the obtained equations of motion, the Lagrange
multiplier function is obtained,
\begin{equation}\label{eq:lambda}
    \lambda = \frac{2\dot{H}}{\mu^4\kappa^2} + 8H^2h''(\chi) + \frac{8}{\mu^2}\left(2\dot{H} - H^2\right)Hh'(\chi).
\end{equation}
Furthermore, we use Planck units, with $H_0 = 1$ together with
natural units $c = \hbar = 1$ and $\kappa^2 = 8\pi G = 1$. While
the original work \cite{Nojiri:2019dwl} focused on demonstrating
the theoretical viability of the model through phenomenological
parameter choices, the present study extends this analysis by
performing a systematic confrontation with real CMB data. The
dimensionless normalization ensures consistency across all
coupling functions $h(\chi)$ and provides a framework for the
Bayesian MCMC parameter estimation carried out in the following
sections.

\subsection{Inflationary Observables}

The ghost-free $f(R,\mathcal{G})$ model represents a special case
of the Einstein-scalar-Gauss-Bonnet model, for which an extended
set of slow-roll parameters based on the cosmological perturbation
theory has been developed in \cite{Hwang:2005hb}. Following this
formalism, all inflationary parameters are expressed in terms of
the e-folds number,
\begin{equation}
    N = \int_{t_i}^{t_f} H(t) dt,
\end{equation}
where $t_i$ and $t_f$ are the initial and final moments of
inflation respectively. Taking into account the features of the
model, the slow-roll parameters are rewritten in terms of e-folds
as follows \cite{Hwang:2005hb},
\begin{align}\label{eq:slow-roll}
    \epsilon_1 &= \frac{H'(N)}{H(N)}, \quad
    \epsilon_2 = \epsilon_3 = 0, \quad
    \epsilon_4 = \frac{1}{2}\frac{E'(N)}{E(N)}, \nonumber \\
    \epsilon_5 &= \frac{4H(N)h'(N)}{1 + 8H^2(N)h'(N)}, \quad
    \epsilon_6 = \frac{H(N)\left(16H'(N)h'(N)
    + 8H(N)h''(N)\right)}{1 + 8H^2(N)h'(N)},
\end{align}
where $\epsilon_2 = \epsilon_3 = 0$ follows directly from $\chi =
\mu^2 t$ and $F = 1$ in the present model. In this case, the
function $E(N)$ is an effective kinetic term and is calculated
using the formula
\begin{equation}
    E(N) = -\lambda H(N)\chi'(N) + \frac{96H^5(N)\left[h'(N)\right]^2}
    {\chi'(N)\left(1 + 8H^2(N)h'(N)\right)},
\end{equation}
where $\lambda$ is the Lagrange multiplier given by
Eq.~(\ref{eq:lambda}). As a result, using the slow-roll parameters
(\ref{eq:slow-roll}), it becomes possible to calculate the
spectral index of scalar perturbations $n_s(N)$ and the
tensor-scalar ratio $r(N)$,
\begin{equation}\label{eq:nsr}
    n_s = 1 + 2\,\frac{\epsilon_1 - \epsilon_4}{1 + \epsilon_1}, \quad
    r = 4\left|\left[\epsilon_1 +
    \frac{Q_e}{4H} - \frac{Q_f}{4}\right]
    \frac{2}{(2 + Q_b)\,c_T^{3}}
    \right|,
\end{equation}
where the auxiliary functions $Q_b$, $Q_e$, $Q_f$ and the tensor
perturbations speed $c_T$ are defined as \cite{Hwang:2005hb},
\begin{align}
    Q_b &= 16H^2 h'(N), \quad Q_e = 32H^2 H'(N) h'(N), \nonumber \\
    Q_f &= -16\left[H^2 h''(N) + HH'(N)h'(N) - H^2h'(N)\right],\\
    c_T^2 &= 1 + \frac{Q_f}{2+Q_b}.
\end{align}
The spectral index of scalar perturbations $n_s$ characterizes the
scale dependence of the primordial power spectrum. Observations
indicate that the spectrum has a red tilt ($n_s<1$), which
indicates power suppression on small scales.

\subsection{Coupling Functions and Hubble Parametrizations}

\begin{figure}[hb]
    \centering
    \includegraphics[width=0.4\linewidth]{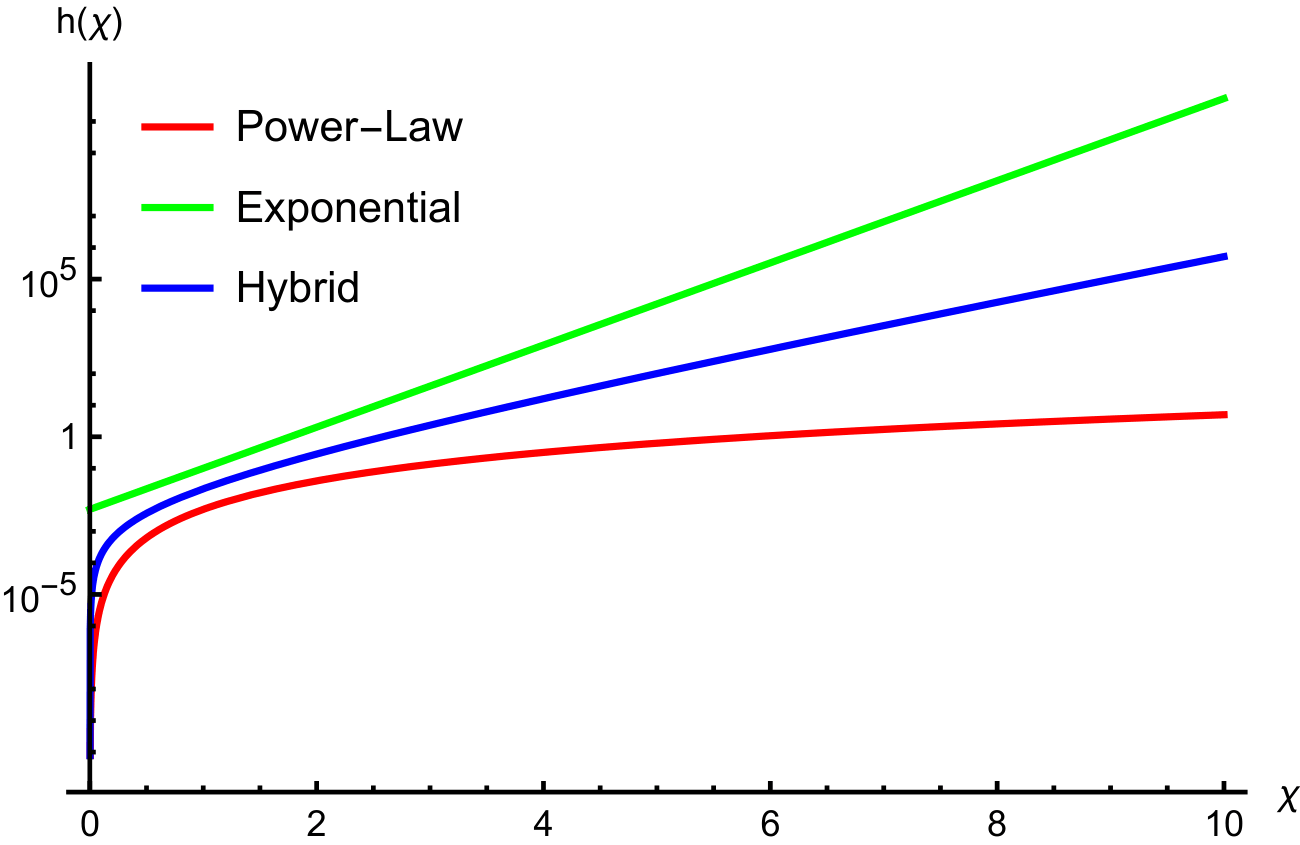}
\caption{Coupling function $h(\chi)$ for power-law (red): $h(\chi)
= \gamma\chi^b$; exponential (green): $h(\chi)=\gamma e^{b\chi}$
and hybrid (blue): $h(\chi)=\gamma e^{b_1 \chi} \chi^{b_2}$ with
$b_1=b_2$ forms.}
    \label{Potentials}
\end{figure}

In this paper, four types of Hubble parameter evolution are
considered, covering a wide class of possible inflationary
scenarios. The De Sitter background ($H= H_0$) is the simplest
analytical case and serves to calibrate the method. The quasi-de
Sitter background ($H = H_0 - H_1 t$) is a more realistic
approximation in which a small correction of $H_1 t$ changes de
Sitter dynamics. Exponential evolution ($H=H_0\mathrm{e}^{-\Omega
t}$) at small values of $\Omega$ asymptotically approaches the de
Sitter regime, but provides a natural way out of inflation at
finite time $t_f =\Omega^{-1}\ln(H_0/\Omega)$. The fractional form
($H=n/t$) is inspired by power-law inflation with a scale factor
$a(t)\propto t^n$ and is a classic parametrization in studies of
the inflationary stage. Each of the listed scenarios is combined
with four types of coupling function $h(\chi)$, which gives
sixteen models, systematized in the Table~\ref{tab:models}.
\begin{table}[ht]
\centering
\caption{Summary of the sixteen inflationary models (labels as in the figures)}
\label{tab:models}
\begin{tabular}{lllll}
\hline\hline
$H(t)$ & Power-law & Exponential & Hybrid & Inv. logarithmic \\
\hline
$H_0$ &
    \textbf{1.1} $\gamma\chi^b$ &
    \textbf{1.2} $\gamma\mathrm{e}^{b\chi}$ &
    \textbf{1.3} $\gamma\mathrm{e}^{b_1\chi}\chi^{b_2}$ &
    \textbf{1.4} $\gamma/\ln(\chi/b)$ \\
$H_0 - H_1 t$ \qquad \qquad&
    \textbf{2.1} $\gamma\chi^b$ &
    \textbf{2.2} $\gamma\mathrm{e}^{b\chi}$ &
    \textbf{2.3} $\gamma\mathrm{e}^{b_1\chi}\chi^{b_2}$ &
    \textbf{2.4} $\gamma/\ln(\chi/b)$ \\
$H_0\mathrm{e}^{-\Omega t}$ &
    \textbf{3.1} $(\chi/M)^n$ \quad&
    \textbf{3.2} $\mathrm{e}^{-\alpha\chi}$ &
    \textbf{3.3} $\mathrm{e}^{-\alpha\chi}\chi^n$ &
    \textbf{3.4} $\gamma/\ln(\chi/b)$ \\
$n/t$ &
    \textbf{4.1} $\gamma\chi^b$ &
    \textbf{4.2} $\gamma\mathrm{e}^{-b\chi}$ \qquad \qquad&
    \textbf{4.3} $\gamma\mathrm{e}^{-b_1\chi}\chi^{b_2}$ \quad&
    \textbf{4.4} $\gamma/\ln(\chi/b)$ \\
\hline\hline
\end{tabular}
\end{table}
Power-law coupling function $h{\chi} =\gamma\chi^b$ represents a
base case with rapid growth at the beginning and continued
moderate growth. The exponential type $h(\chi) =
\gamma\mathrm{e}^{b\chi}$, on the contrary, is characterized by a
faster increase and provides an alternative mode of contribution
of the Gauss-Bonnet invariant. The developed hybrid form $h(\chi)=
\gamma\mathrm{e}^{b_1\chi}\chi^{b_2}$ combines an early rapid
growth of the power-law type and a late increase of the
exponential type (Fig.~\ref{Potentials}), which provides
additional flexibility in adjusting the contribution of the
Gauss-Bonnet invariant. The direct logarithmic coupling
$h(\chi)=\gamma\ln(\chi/b)$ demonstrates a blue tilt of the
spectrum ($n_s>1$). For this reason, the inverse logarithmic
coupling $h(\chi) =\gamma/\ln(\chi/b)$ is used, which demonstrates
viability and compliance with observational data.

\section{Method}\label{sec:method}

The calculations are performed using the code for Bayesian MCMC
analysis \texttt{Cobaya}. For each of the sixteen models, the most
probable parameter values were calculated and used to compute the
spectral index of scalar perturbations $n_s$ and the tensor-scalar
ratio $r$ at $N=60$ e-folds. First, preliminary calculations were
carried out aimed at determining the probability of blue or red
spectral tilts in a wide range of values of $n_s \in [0,2]$.
Further, after identifying viable models, calculations were
performed on Planck and ACT datasets to demonstrate the viability
of the models and determine preferred parameter values.

The Planck 2018 low-$\ell$ TT and EE likelihoods break the
$A_s$-$\tau$ degeneracy, enabling reliable extraction of the
spectral index $n_s$ \cite{Planck:2018jri}. The ACT DR6 CMB-only
TT+TE+EE likelihood, with foregrounds pre-marginalised, provides
the primary high-$\ell$ constraints on the shape of the primordial
power spectrum and hence on $n_s$ \cite{ACT:2025fju, ACT:2025tim}.
The BICEP/Keck~2018 B-mode likelihood is the sole direct source of
constraints on the tensor-to-scalar ratio $r$
\cite{BICEP:2021xfz}.\footnote{Cobaya likelihoods:
\texttt{planck\_2018\_lowl.TT}, \texttt{planck\_2018\_lowl.EE},
\texttt{act\_dr6\_cmbonly.ACTDR6CMBonly},
\texttt{bicep\_keck\_2018}.} All computations use \texttt{CAMB}
\cite{Lewis:1999bs} as the Boltzmann solver.

For visual comparison with the observational data, the results of
model calculations are displayed together with the contours of the
datasets on the $\{ns,r\}$ plane. The reference contours are taken
from the official releases of the chains of relevant
collaborations. The contours of Planck 2018 are obtained from the
Planck Legacy Archive\footnote{Configuration of Planck 2018
reference contours:
\texttt{base\_r\_CamSpecHM\_TTTEEE\_lowl\_lowE\_BK15\_post\_BAO\_lensing}}.
The contours of ACT DR6.02 are obtained from the NASA LAMBDA
archive\footnote{Configuration of ACT DR6.02 reference contours:
\texttt{actlite-bk-l-b\_r\_camb}}.

The parameter $\mu$ defines the scale of the auxiliary field $\chi
= \mu^2 t$ of the entire ghost-free $f(R,\mathcal{G})$ model,
therefore it is globally limited for all cases. The remaining
parameters are determined individually depending on the results of
the preliminary calculations. Chain convergence is monitored using
the Gelman-Rubin criterion, with sampling automatically terminated
once $R-1 < 0.01$ is reached. The calculation of inflationary
parameters requires multiple numerical differentiation of the
functions $h(\chi)$ and $H(N)$. To reduce computational costs,
numerical differentiation with a fixed number of points is
employed, which leads to minor instabilities on $n_s(N)$ and
$r(N)$ curves. However, this effect does not affect the accuracy
of the results, since the MCMC analysis requires only a value at
the point $N=60$.

\section{Results: de Sitter Background}\label{sec:dS}

\subsection{Power-Law Coupling}

The power-law coupling function $h(\chi) = \gamma\chi^b$ is the
base case for de Sitter background $H=H_0$. The dynamics of the
spectral index of scalar perturbations $n_s(N)$ demonstrates
monotonous growth in the inflationary range of e-folds $50\leq
N\leq 60$. A similar growth dynamics is observed for the
tensor-scalar ratio $r(N)$. The best values of the model
parameters at $N = 60$ are $n_s = 0.9686^{+0.0038}_{-0.0045}$ and
$r = 0.0025^{+0.0020}_{-0.0015}$.
\begin{figure}[ht]
\centering
\includegraphics[width=0.96\linewidth]{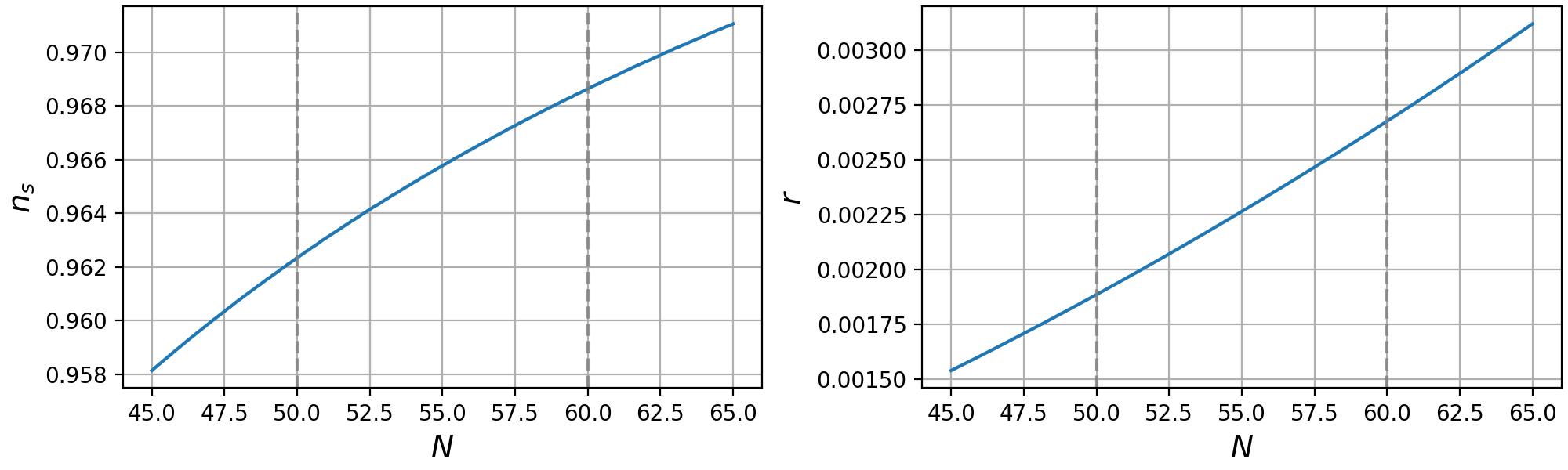}
\begin{minipage}{0.48\linewidth}
    \centering
    \includegraphics[width=\linewidth]{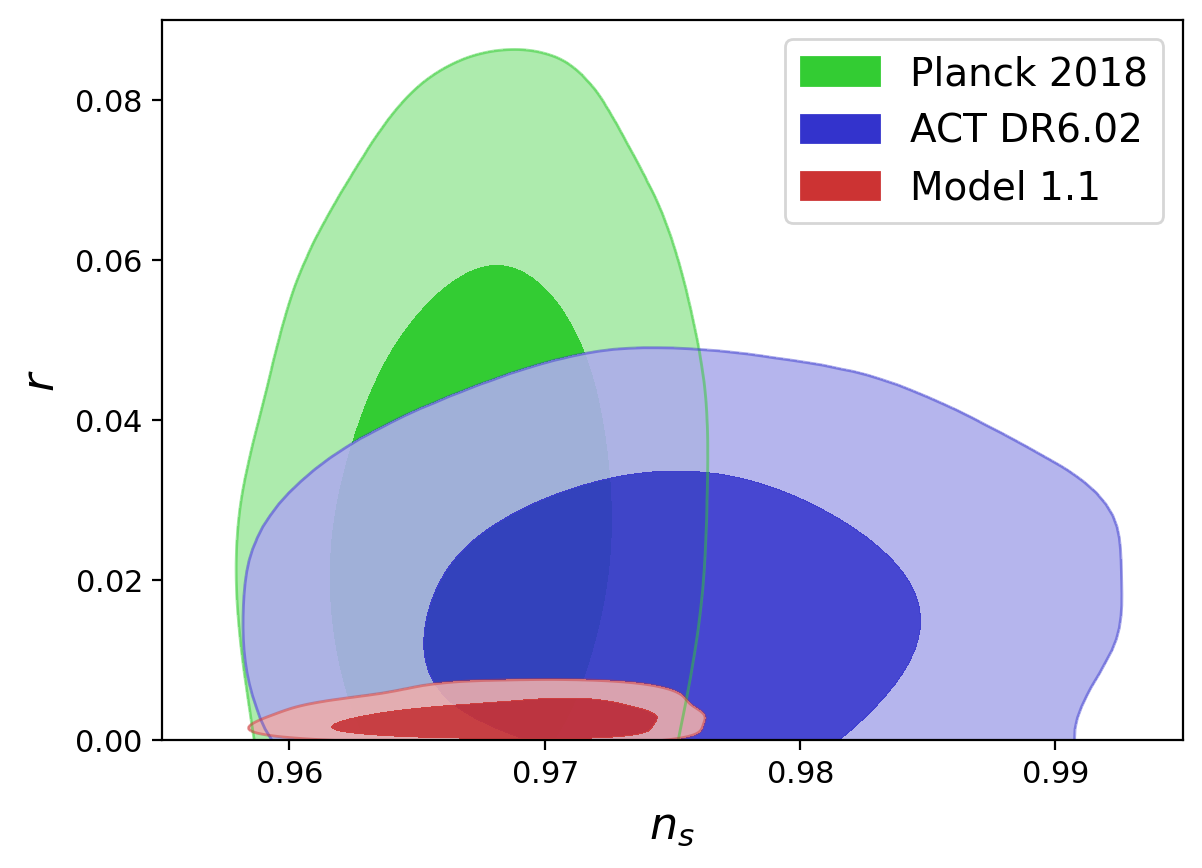}
\end{minipage}
\hfill
\begin{minipage}{0.48\linewidth}
    \centering
    \renewcommand{\arraystretch}{1.7}
    \begin{tabular}{cc}
    \toprule
    Parameter & Value \\
    \midrule
    $\mu$    & $0.0997^{+0.0061}_{-0.0063}$ \\
    $\gamma$ & $0.0159^{+0.0088}_{-0.0092}$ \\
    $b$      & $2.8819^{+0.2684}_{-0.2289}$ \\
    $n_s$    & $0.9686^{+0.0038}_{-0.0045}$ \\
    $r$      & $0.0025^{+0.0020}_{-0.0015}$ \\
    \bottomrule
    \space & \space
    \end{tabular}
\end{minipage}
\caption{Model~1.1: power-law coupling $h(\chi) = \gamma\chi^b$
on the de~Sitter background $H = H_0$. The dynamic plots show
the evolution of $n_s(N)$ (left) and $r(N)$ (right) over the
inflationary window $50 \leq N \leq 60$. The contour plot shows
marginalized constraints on the $(n_s,\, r)$ plane at $N = 60$
against Planck~2018 (green) and ACT~DR6.02 (blue) at $1\sigma$
and $2\sigma$ confidence levels. The table lists best-fit
parameter values $\pm1\sigma$.}
\label{fig:11combined}
\end{figure}
The contour plot indicates a clear preference for Planck 2018
data, as the $1\sigma$ area of the model is completely covered by
the Planck values. The contour of the model practically does not
overlap with the central contour of the ACT, but remains within
the $2\sigma$ areas of both datasets, which indicates the
viability of the model. The contour shape is close to oval with
clearly distinguishable $1\sigma$ and $2\sigma$ regions,
indicating good MCMC convergence and well-constrained parameter
posteriors. Among all four coupling functions considered on the
de~Sitter background, the power-law form yields the best overall
fit quality and the most compact posterior distribution. The value
of $\mu = 0.0997^{+0.0061}_{-0.0063}$ reproduces the
characteristic value of $\mu\approx 0.10$, which is observed in
all models.

\subsection{Exponential Coupling}

The exponential coupling function $h(\chi) =
\gamma\mathrm{e}^{b\chi}$ is characterized by faster growth
compared to the power-law case. The dynamics of $n_s(N)$ in the
inflationary window $50\leq N\leq 60$ demonstrates a weak
dependence on the number of e-folds. The spectral index changes
only in the fourth decimal place, ranging from $n_s\approx 0.9702$
at $N = 50$ to $n_s\approx 0.9703$ at $N = 60$. Such a small range
of changes is due to the peculiarities of the exponential coupling
function on a de Sitter background. With a constant Hubble
parameter $H = H_0$, the contribution of the Gauss-Bonnet
invariant is exponentially saturated, which leads to stabilization
of inflationary parameters. On the contrary, the tensor-to-scalar
ratio $r(N)$ demonstrates a wider range of values. The best values
of the model parameters at $N = 60$ are $n_s =
0.9703^{+0.0047}_{-0.0044}$ and $r = 0.0146^{+0.0128}_{-0.0080}$.
\begin{figure}[ht]
\centering
\includegraphics[width=0.96\linewidth]{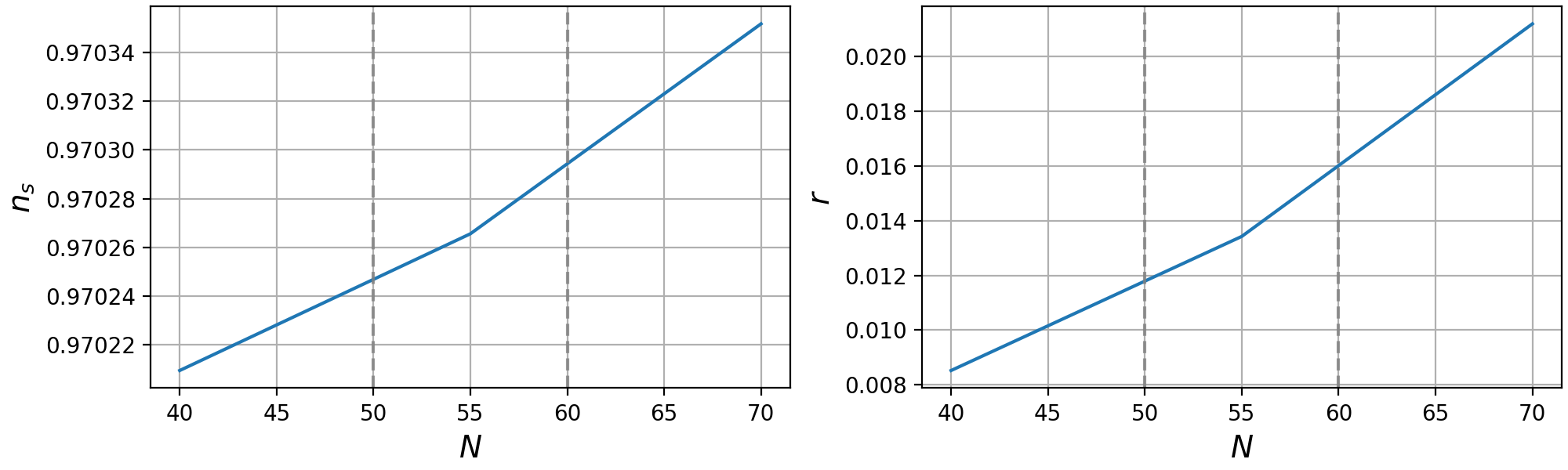}
\begin{minipage}{0.48\linewidth}
    \centering
    \includegraphics[width=\linewidth]{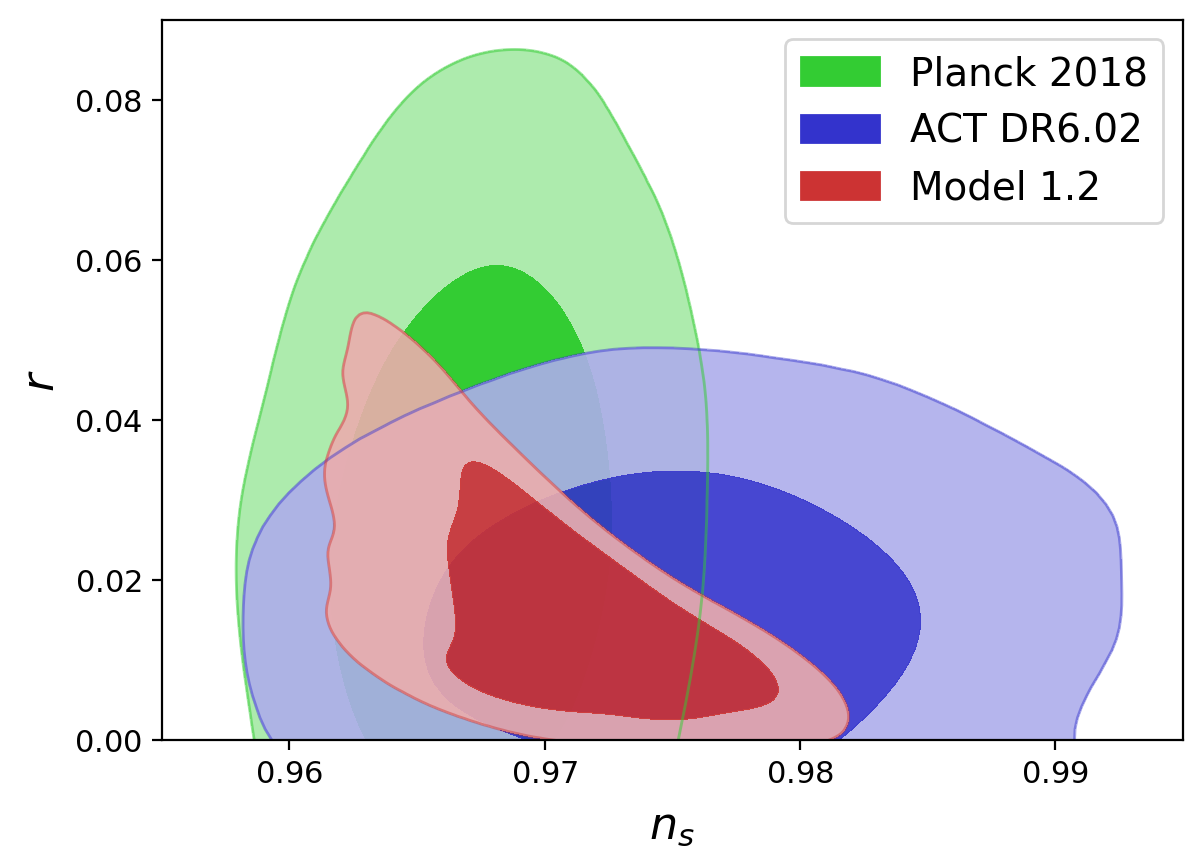}
\end{minipage}
\hfill
\begin{minipage}{0.48\linewidth}
    \centering
    \renewcommand{\arraystretch}{1.7}
    \begin{tabular}{cc}
    \toprule
    Parameter & Value \\
    \midrule
    $\mu$    & $0.0997^{+0.0061}_{-0.0058}$ \\
    $\gamma$ & $0.0055^{+0.0028}_{-0.0028}$ \\
    $b$      & $3.0107^{+0.5417}_{-0.5212}$ \\
    $n_s$    & $0.9703^{+0.0047}_{-0.0044}$ \\
    $r$      & $0.0146^{+0.0128}_{-0.0080}$ \\
    \bottomrule
    \space & \space
    \end{tabular}
\end{minipage}
\caption{Model~1.2: exponential coupling
$h(\chi) = \gamma\mathrm{e}^{b\chi}$ on the de~Sitter background
$H = H_0$. The dynamic plots show the evolution of $n_s(N)$
(left) and $r(N)$ (right) over the inflationary window
$50 \leq N \leq 60$. The contour plot shows marginalized
constraints on the $(n_s,\, r)$ plane at $N = 60$ against
Planck~2018 (green) and ACT~DR6.02 (blue) at $1\sigma$ and
$2\sigma$ confidence levels. The table lists best-fit parameter
values $\pm1\sigma$.}
\label{fig:12combined}
\end{figure}
The $n_s(N)$ and $r(N)$ graphs show a characteristic instability
in the form of stepped curves. As noted in
section~\ref{sec:method}, this effect is a consequence of
discretized numerical differentiation and is especially noticeable
for small ranges of function variation. In the case of an
exponential relationship, $n_s$ changes only at the fourth decimal
place, which increases the visibility of this artifact. However,
this does not affect the accuracy of the results, since the MCMC
analysis is based solely on the values observed at $N = 60$.

The contour graph shows that the model shows a preference for both
datasets approximately equally. The $1\sigma$ region of the model
overlaps with both the $2\sigma$ Planck region and the $2\sigma$
ACT region, which indicates the viability of the model in the
presence of some tension with both datasets. The contour shape is
somewhat slanted diagonally and differs from the ideal oval shape,
which indicates an asymmetry in the a posteriori distribution of
parameters. In particular, the parameters $\gamma$ and $b$ exhibit
correlated behavior. As $b$ increases, the exponential function
grows faster, which is offset by a decrease in the amplitude of
$\gamma$ to preserve the required contribution of the Gauss-Bonnet
invariant. This correlation leads to the characteristic obliquity
of the contour.

Compared to the 1.1 model, the exponential coupling gives a
noticeably higher tensor-to-scalar ratio for similar spectral
index values. This is explained by the fact that the exponential
potential provides a more intensive contribution of tensor
perturbations through enhanced interaction with the Gauss-Bonnet
invariant. The value of $r = 0.0146^{+0.0128}_{-0.0080}$ remains
below the current observational limit $r < 0.036$. The parameter
$\mu = 0.0997^{+0.0061}_{-0.0058}$ reproduces the characteristic
value of $\mu\approx 0.10$.

\subsection{Hybrid Coupling}

The hybrid coupling function $h(\chi) =
\gamma\mathrm{e}^{b_1\chi}\chi^{b_2}$ is introduced in order to
combine the advantages of the power-law and exponential forms of
coupling potentials. As shown in Fig.~\ref{Potentials}, a
power-law potential is characterized by rapid growth at the
beginning. The exponential potential dominates in the later
stages. Precise control of the rate of increase of the function
$h(\chi)$ is necessary to adjust the effect of the Gauss-Bonnet
invariant. The classical EsGB model uses a scalar field $\phi$,
whose function decreases, approaching zero in the late Universe.
Therefore, the potential $\xi(\phi)$ also decreases, thereby
reducing the effect of the Gauss-Bonnet invariant in later times.
However, the ghost-free model uses the auxiliary field $\chi=\mu^2
t$, which grows linearly over time. Because of this, the function
$h(\chi)$ is also growing, which increases the duration of the
influence of the Gauss-Bonnet invariant on the dynamics of the
Universe.

Therefore, it is necessary to achieve a balance in which the
function $h(\chi)$ does not lead to a large effect of the
Gauss-Bonnet invariant in the late Universe. At the same time, the
power-law potential may grow too slowly and offset the effect of
the Gauss-Bonnet invariant in the early stages. Therefore, the
introduction of a hybrid potential makes it possible to expand the
possibilities in matching the inflationary and late expansion
dynamics.
\begin{figure}[ht]
\centering
\includegraphics[width=0.96\linewidth]{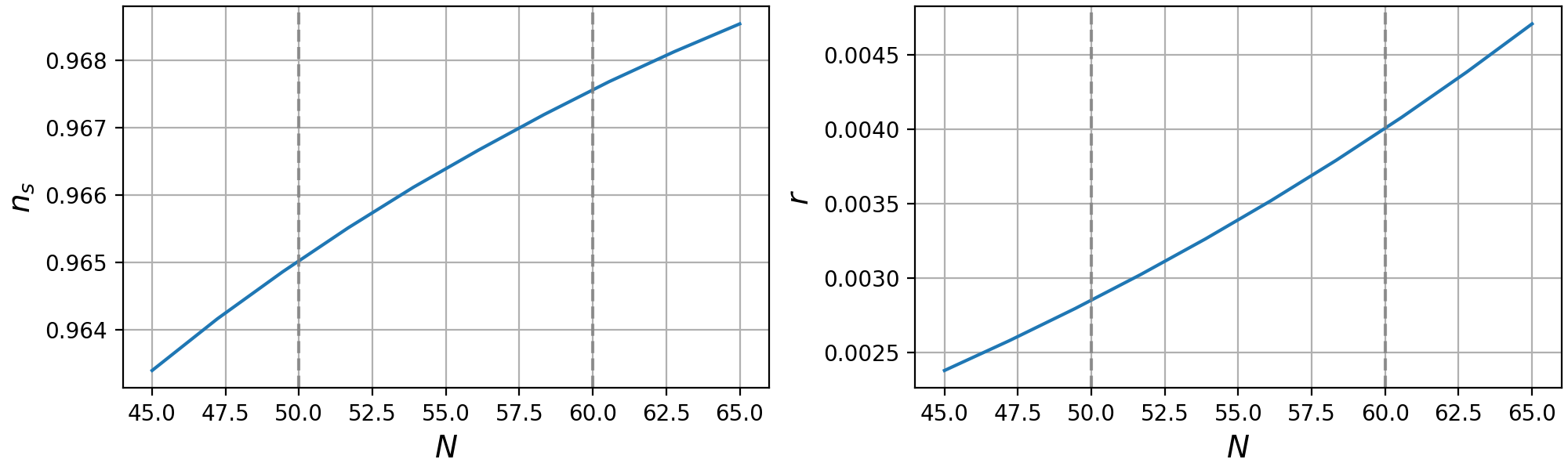}
\begin{minipage}{0.48\linewidth}
    \centering
    \includegraphics[width=\linewidth]{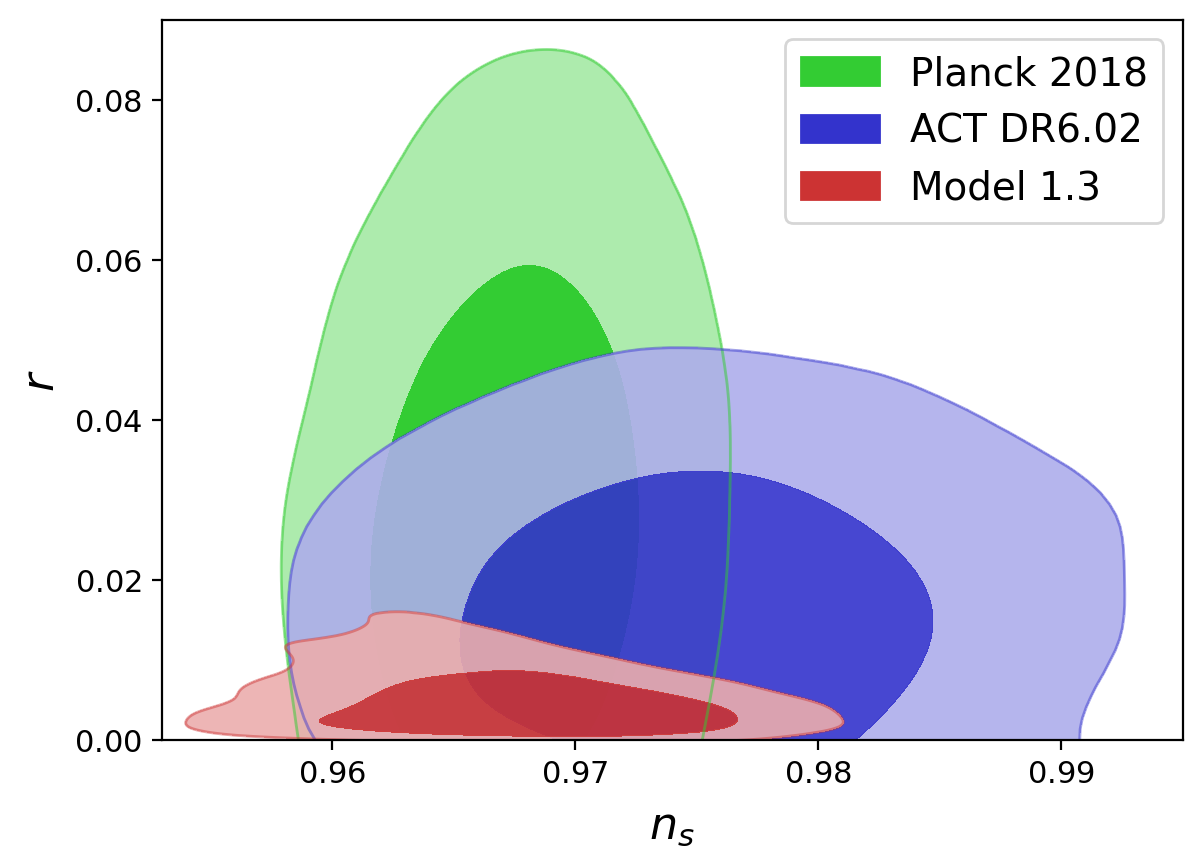}
\end{minipage}
\hfill
\begin{minipage}{0.48\linewidth}
    \centering
    \renewcommand{\arraystretch}{1.7}
    \begin{tabular}{cc}
    \toprule
    Parameter & Value \\
    \midrule
    $\mu$    & $0.0985^{+0.0067}_{-0.0057}$ \\
    $\gamma$ & $0.0056^{+0.0028}_{-0.0029}$ \\
    $b_1$    & $1.6326^{+0.5370}_{-0.4106}$ \\
    $b_2$    & $1.5409^{+0.4362}_{-0.3549}$ \\
    $n_s$    & $0.9675^{+0.0051}_{-0.0052}$ \\
    $r$      & $0.0039^{+0.0036}_{-0.0021}$ \\
    \bottomrule
    \space & \space
    \end{tabular}
\end{minipage}
\caption{Model~1.3: hybrid coupling
$h(\chi) = \gamma\mathrm{e}^{b_1\chi}\chi^{b_2}$ on the de~Sitter
background $H = H_0$. The dynamic plots show the evolution of
$n_s(N)$ (left) and $r(N)$ (right) over the inflationary window
$50 \leq N \leq 60$. The contour plot shows marginalized
constraints on the $(n_s,\, r)$ plane at $N = 60$ against
Planck~2018 (green) and ACT~DR6.02 (blue) at $1\sigma$ and
$2\sigma$ confidence levels. The table lists best-fit parameter
values $\pm1\sigma$.}
\label{fig:13combined}
\end{figure}

The dynamics of $n_s(N)$ demonstrates monotonous growth in the
inflationary e-fold range $50\leq N\leq 60$, as well as the
tensor-to-scalar ratio $r(N)$. The best parameter values for $N =
60$ are $n_s= 0.9675^{+0.0051}_{-0.0052}$ and $r =
0.0039^{+0.0036}_{-0.0021}$, preferring mainly the data from
Planck 2018.

Values of the parameter $\mu = 0.0985^{+0.0067}_{-0.0057} \approx
0.1$ preserved in this model. In comparison with the contour of
the exponential coupling potential (Fig.~\ref{fig:12combined}),
the contour of the hybrid case is more compact and has a more
uniform distribution. This contour shape indicates the best
quality of the selection of parameters. However, in comparison
with the contour of the power-law coupling
(Fig.~\ref{fig:11combined}), the distribution is less uniform. It
can be seen that the contour is a combination of two previous
types and is located in the confidence interval $2\sigma$ of both
datasets, thereby demonstrating the viability of the model. The
values of the balancing parameters $b_1 =
1.6326^{+0.5370}_{-0.4106}$ and $b_2 = 1.5409^{+0.4362}_{-0.3549}$
have similar values, but there is a slight dominance of the
exponential coupling.

\subsection{Inverse Logarithmic Coupling}

\begin{figure}[htb]
    \centering
    \includegraphics[width=0.96\linewidth]{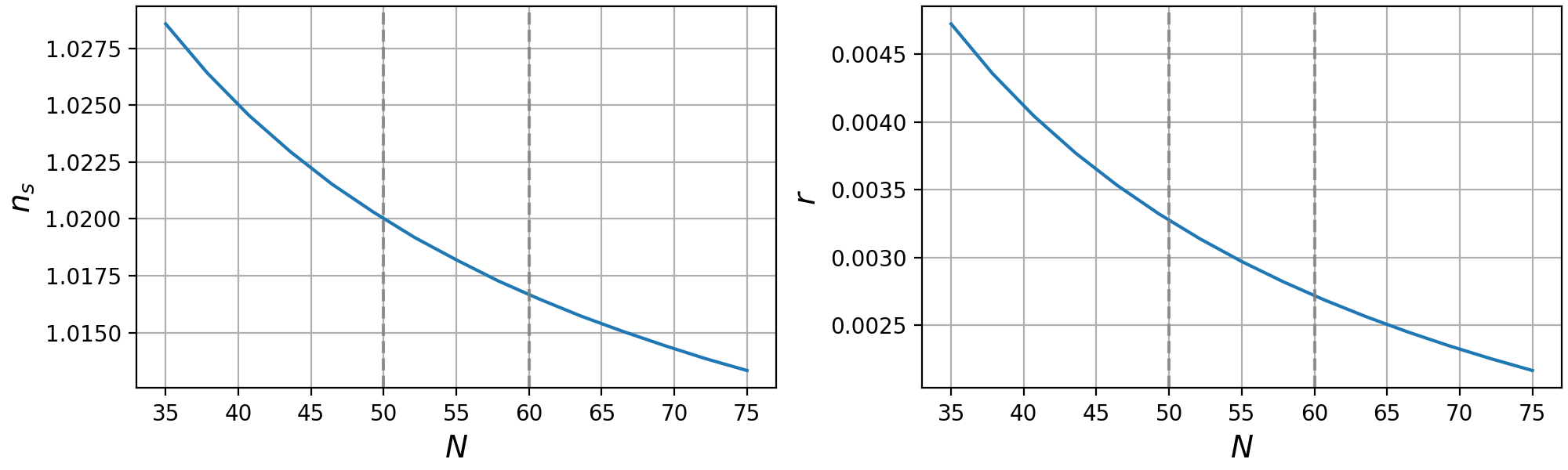}
    \caption{The dynamics of the spectral index of scalar perturbations
$n_s(N)$ and the tensor-to-scalar ratio $r(N)$ in the inflationary
range $50\leq N\leq 60$ with a direct logarithmic coupling
$h(\chi) = \gamma\ln(\chi/b)$.}
    \label{fig:140nsr}
\end{figure}

\begin{figure}[ht]
\centering
\includegraphics[width=0.96\linewidth]{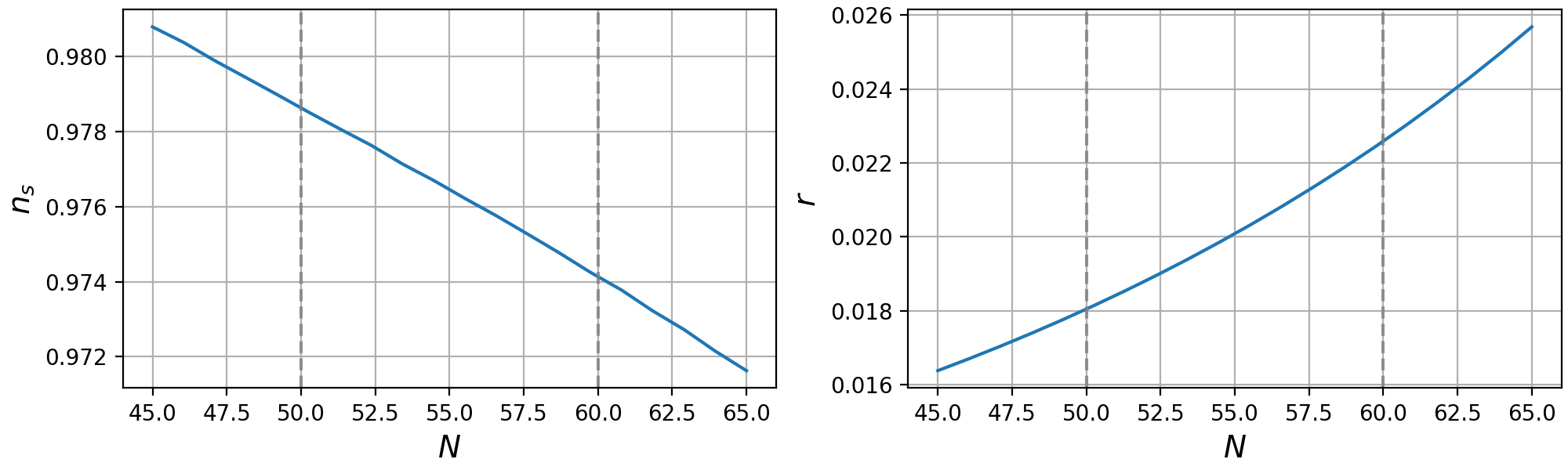}
\begin{minipage}{0.48\linewidth}
    \centering
    \includegraphics[width=\linewidth]{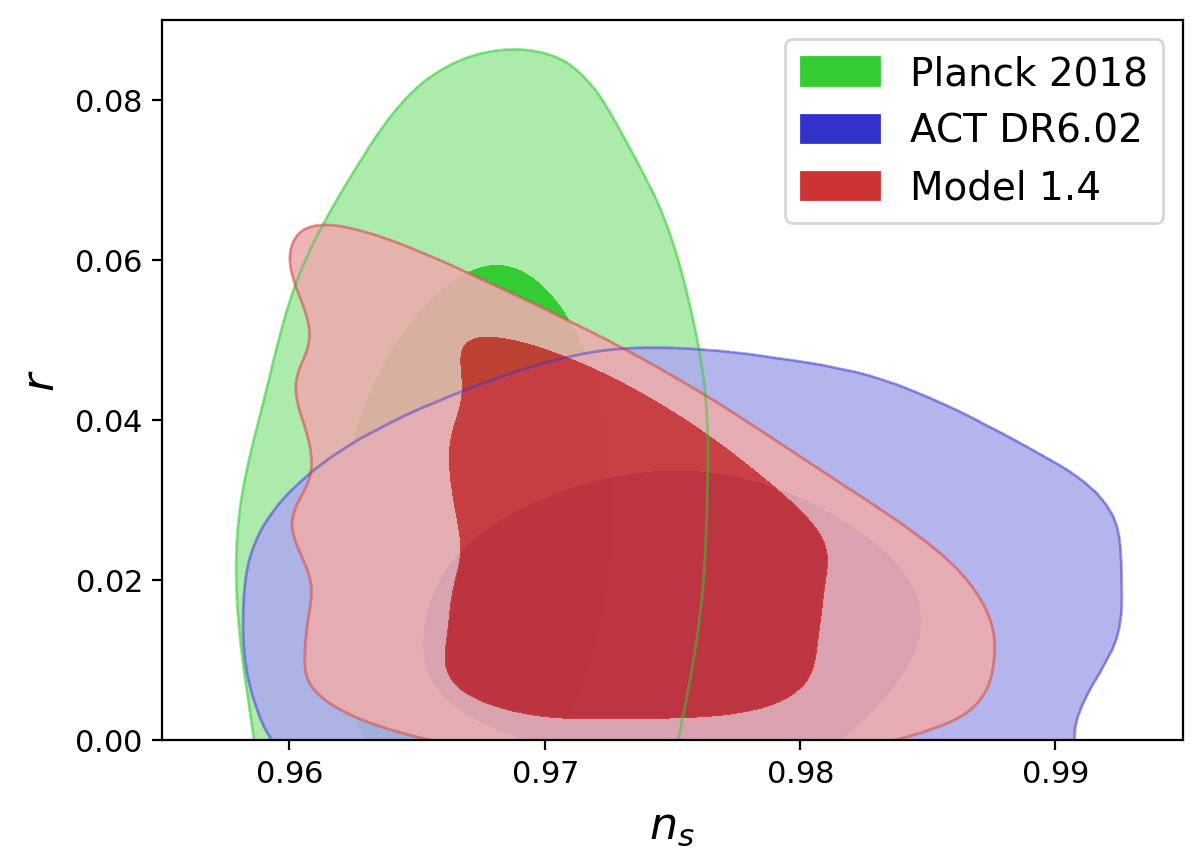}
\end{minipage}
\hfill
\begin{minipage}{0.48\linewidth}
    \centering
    \renewcommand{\arraystretch}{1.7}
    \begin{tabular}{cc}
    \toprule
    Parameter & Value \\
    \midrule
    $\mu$    & $0.1011^{+0.0058}_{-0.0067}$ \\
    $\gamma$ & $0.0554^{+0.0339}_{-0.0341}$ \\
    $b$      & $1.3200^{+0.2255}_{-0.1933}$ \\
    $n_s$    & $0.9729^{+0.0059}_{-0.0064}$ \\
    $r$      & $0.0228^{+0.0157}_{-0.0140}$ \\
    \bottomrule
    \space & \space
    \end{tabular}
\end{minipage}
\caption{Model~1.4: inverse logarithmic coupling
$h(\chi) = \gamma/\ln(\chi/b)$ on the de~Sitter background
$H = H_0$. The dynamic plots show the evolution of $n_s(N)$
(left) and $r(N)$ (right) over the inflationary window
$50 \leq N \leq 60$. The contour plot shows marginalized
constraints on the $(n_s,\, r)$ plane at $N = 60$ against
Planck~2018 (green) and ACT~DR6.02 (blue) at $1\sigma$ and
$2\sigma$ confidence levels. The table lists best-fit parameter
values $\pm1\sigma$.}
\label{fig:14combined}
\end{figure}

Initially, it was assumed that the direct logarithmic relationship
$h(\chi) = \gamma\ln(\chi/b)$ would be viable. However,
preliminary calculations have shown that for any parameter values,
this form exhibits a blue tilt of the spectrum ($n_s > 1$), as
shown in Fig.~\ref{fig:140nsr}. Despite the fact that the
tensor-to-scalar ratio has values consistent with observations,
the blue tilt of the spectrum contradicts them. Therefore, a
direct logarithmic coupling is not viable within the framework of
the de Sitter background.
For this reason, the inverse logarithmic coupling $h(\chi) =
\gamma/\ln(\chi/b)$ is investigated, demonstrating compliance with
the observational data, as shown in Fig.~\ref{fig:14combined}.
Unlike previous models, the inverse logarithmic form leads to a
monotonically decreasing spectral index $n_s(N)$. At the same
time, the tensor-to-scalar ratio increases, as in the previous
cases.
The contour plot shows that the model, unlike previous cases,
shows a preference for ACT data. The spectral index $n_s$ is well
constrained, as can be seen by distinguishable $1\sigma$ and
$2\sigma$ regions. However, the tensor-to-scalar ratio is
artificially bounded from above by the prior range of the
parameter $\gamma$. With a wider range of acceptable values for
this parameter, the contour would have the appearance of an
elongated vertical strip rather than a closed region. For this
reason, the inverse logarithmic coupling model has a weaker
predictive power, unlike in the previous three cases. Despite
this, the model is still viable and reproduces the values of $\mu
= 0.1011^{+0.0058}_{-0.0067} \approx 0.10$.

\section{Results: Quasi-de Sitter Background}\label{sec:QdS}

\subsection{Power-Law Coupling}

The power-law coupling function $h(\chi) = \gamma\chi^b$ on a
quasi-de Sitter background $H = H_0 - H_1 t$ differs from the
similar case on a de Sitter background by the presence of an
additional parameter $H_1$. Non-zero values of this parameter set
small linear deviations from the constant expansion mode.
Preliminary calculations have shown that an increase in the
parameter $H_1$ entails a shift in the spectral index of scalar
perturbations $n_s$ towards the blue tilt, up to values of
$n_s>1$. For this reason, the total values of the parameter $H_1$
are quite small ($H_1 = 0.0008^{+0.0009}_{-0.0005}$), but not
zero, which preserves the relevance of its application. The
dynamics of $n_s(N)$ demonstrates monotonous growth, as does the
ratio $r(N)$ in the inflationary range of $50\leq N\leq 60$. The
best values at the end of inflation at $N = 60$ are $n_s =
0.9736^{+0.0049}_{-0.0049}$ and $r = 0.0048^{+0.0051}_{-0.0027}$.
Values of the parameter $\mu = 0.1017^{+0.0056}_{-0.0067}\approx
0.10$.

\begin{figure}[ht]
\centering
\includegraphics[width=0.96\linewidth]{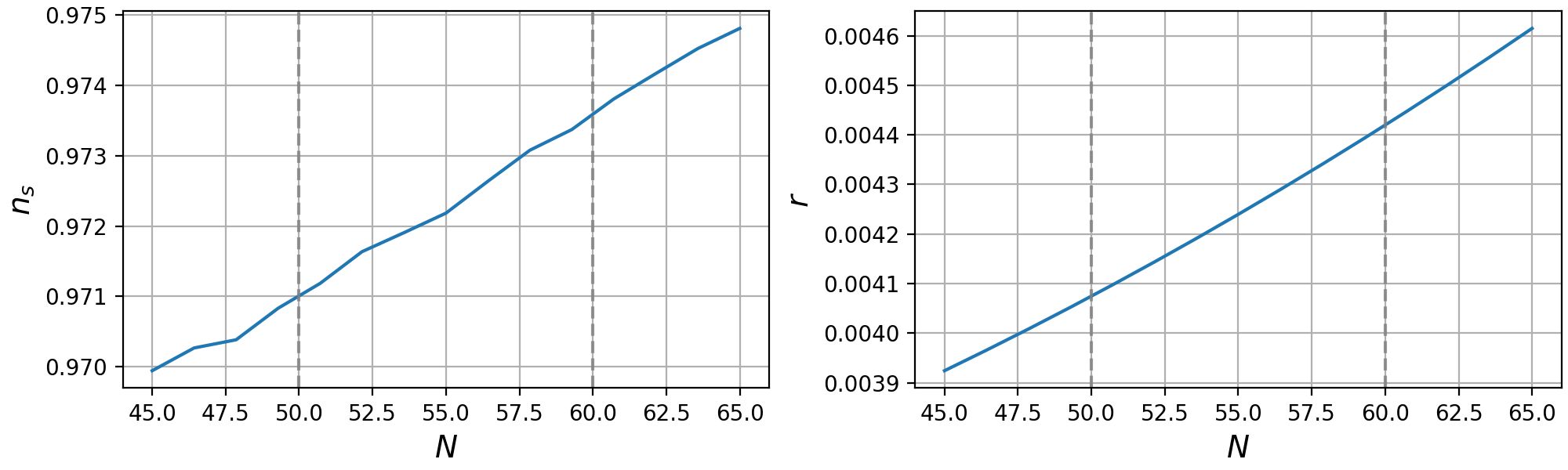}
\begin{minipage}{0.48\linewidth}
    \centering
    \includegraphics[width=\linewidth]{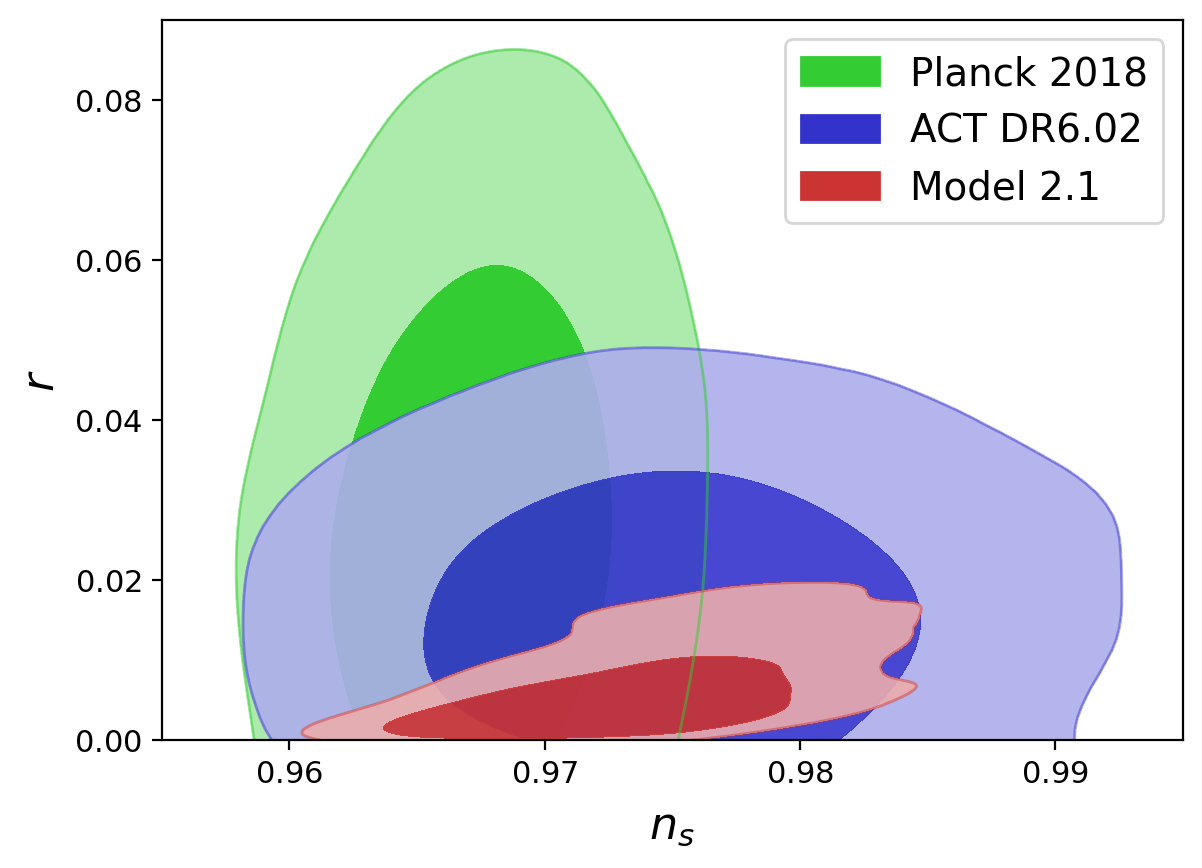}
\end{minipage}
\hfill
\begin{minipage}{0.48\linewidth}
    \centering
    \renewcommand{\arraystretch}{1.7}
    \begin{tabular}{cc}
    \toprule
    Parameter & Value \\
    \midrule
    $H_1$    & $0.0008^{+0.0009}_{-0.0005}$ \\
    $\mu$    & $0.1017^{+0.0056}_{-0.0067}$ \\
    $\gamma$ & $0.0066^{+0.0022}_{-0.0028}$ \\
    $b$      & $3.0966^{+0.2551}_{-0.3178}$ \\
    $n_s$    & $0.9736^{+0.0049}_{-0.0049}$ \\
    $r$      & $0.0048^{+0.0051}_{-0.0027}$ \\
    \bottomrule
    \space & \space
    \end{tabular}
\end{minipage}
\caption{Model~2.1: power-law coupling $h(\chi) = \gamma\chi^b$
on the quasi-de~Sitter background $H = H_0 - H_1 t$. The dynamic
plots show the evolution of $n_s(N)$ (left) and $r(N)$ (right)
over the inflationary window $50 \leq N \leq 60$. The contour
plot shows marginalized constraints on the $(n_s,\, r)$ plane at
$N = 60$ against Planck~2018 (green) and ACT~DR6.02 (blue) at
$1\sigma$ and $2\sigma$ confidence levels. The table lists
best-fit parameter values $\pm1\sigma$.}
\label{fig:21combined}
\end{figure}
Compared with the power-law case in the previous parametrization
$H=H_0$ (Fig.~\ref{fig:11combined}), the model's preference area
has shifted towards the ACT DR6.02 data. The nonzero parameter
$H_1$ simultaneously shifts the spectral index of scalar
perturbations to a bluer region and increases the probable values
of the tensor-to-scalar ratio. Nevertheless, the model remains
viable and is within the $2\sigma$ level of both datasets. The
distribution of the probable parameter values takes on a wing-like
appearance with narrow bright areas (Fig.~\ref{fig:21combined}).
Such a contour indicates a less balanced distribution in
comparison with the similar case on the background of de Sitter.

\subsection{Exponential Coupling}

\begin{figure}[ht]
\centering
\includegraphics[width=0.96\linewidth]{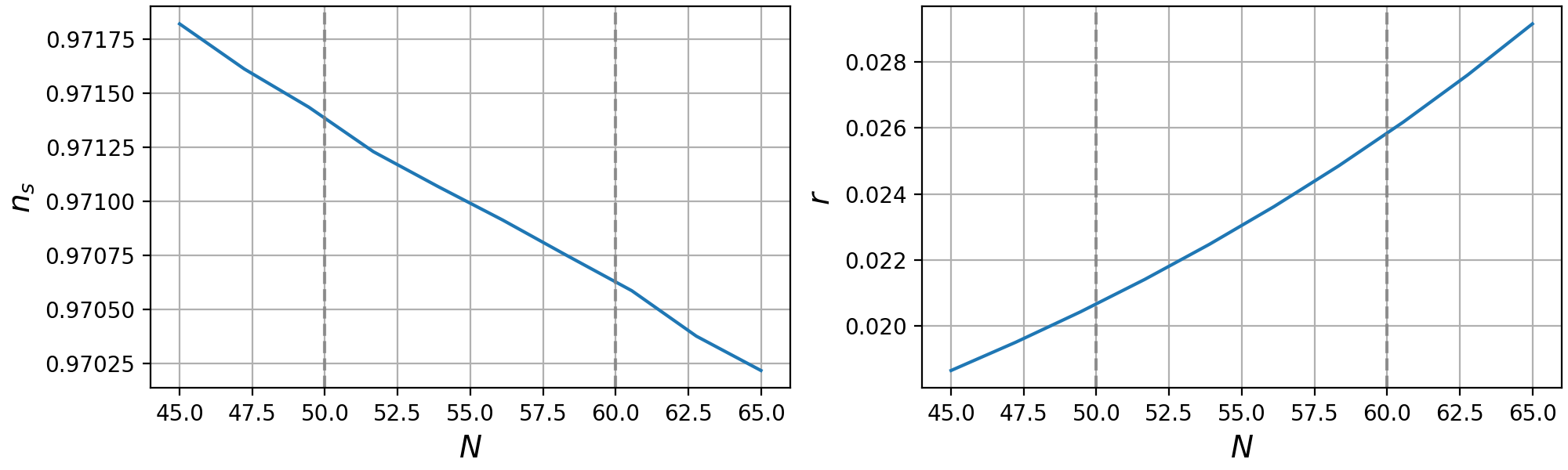}
\begin{minipage}{0.48\linewidth}
    \centering
    \includegraphics[width=\linewidth]{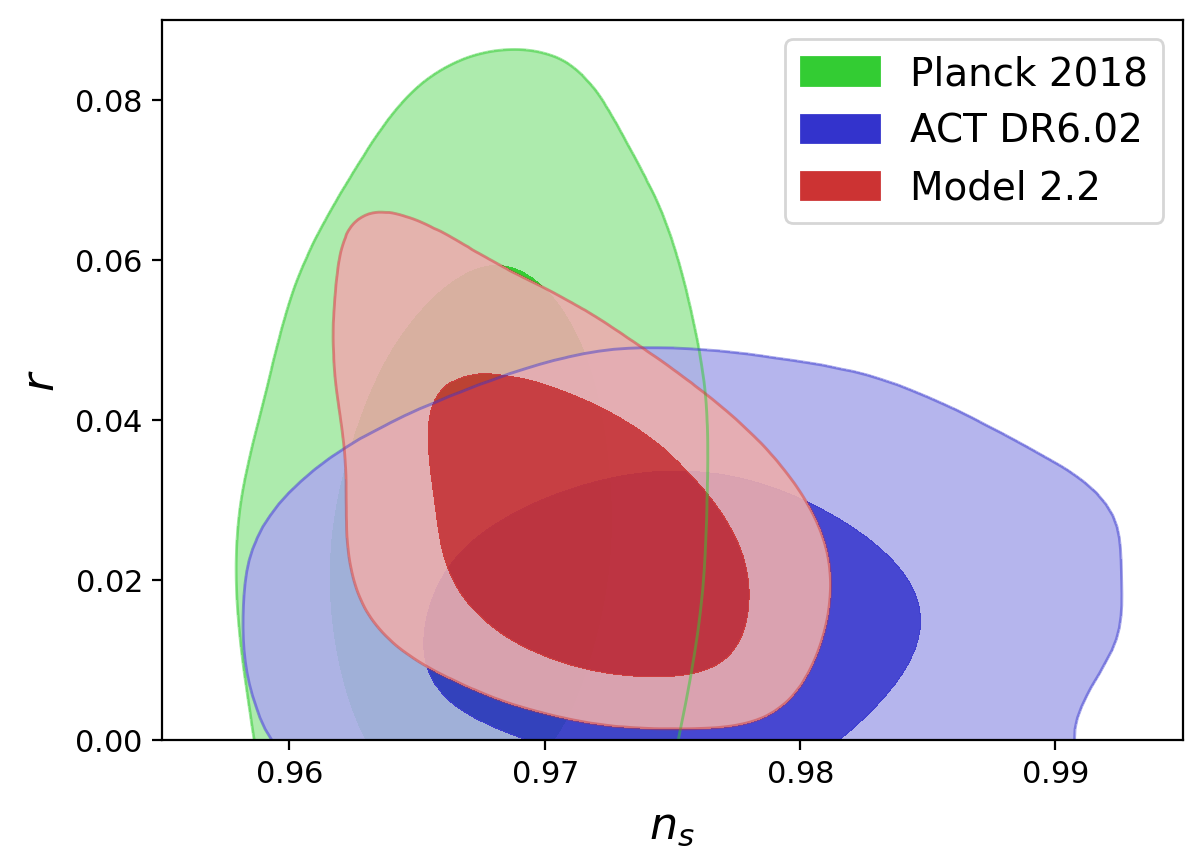}
\end{minipage}
\hfill
\begin{minipage}{0.48\linewidth}
    \centering
    \renewcommand{\arraystretch}{1.6}
    \begin{tabular}{cc}
    \toprule
    Parameter & Value \\
    \midrule
    $H_1$    & $0.0016^{+0.0011}_{-0.0010}$ \\
    $\mu$    & $0.1007^{+0.0057}_{-0.0061}$ \\
    $\gamma$ & $0.0058^{+0.0027}_{-0.0028}$ \\
    $b$      & $3.0541^{+0.2790}_{-0.3248}$ \\
    $n_s$    & $0.9712^{+0.0043}_{-0.0042}$ \\
    $r$      & $0.0257^{+0.0140}_{-0.0108}$ \\
    \bottomrule
    \space & \space
    \end{tabular}
\end{minipage}
\caption{Model~2.2: exponential coupling
$h(\chi) = \gamma\mathrm{e}^{b\chi}$ on the quasi-de~Sitter
background $H = H_0 - H_1 t$. The dynamic plots show the
evolution of $n_s(N)$ (left) and $r(N)$ (right) over the
inflationary window $50 \leq N \leq 60$. The contour plot shows
marginalized constraints on the $(n_s,\, r)$ plane at $N = 60$
against Planck~2018 (green) and ACT~DR6.02 (blue) at $1\sigma$
and $2\sigma$ confidence levels. The table lists best-fit
parameter values $\pm1\sigma$.}
\label{fig:22combined}
\end{figure}

The exponential coupling function $h(\chi) =
\gamma\mathrm{e}^{b\chi}$ against the background of quasi-de
Sitter $H = H_0 - H_1 t$ demonstrates several features that
distinguish it from the similar case against the background of de
Sitter $H=H_0$. The dynamics of the spectral index $n_s$ is
decreasing, while the tensor-to-scalar ratio, on the contrary,
shows growth. A similar behavior was observed in Model 1.4
(Fig.~\ref{fig:14combined}) with an inverse logarithmic coupling
function. Values of the parameter $H_1=0.0016^{+0.0011}_{-0.0010}$
is twice as large as the previous ones described in
Fig.~\ref{fig:21combined}. The increased values of this parameter
indicate that the exponential coupling allows for large deviations
from the constant de Sitter expansion. The best parameter values
for $N = 60$ are $n_s = 0.9712^{+0.0043}_{-0.0042}$ and $r =
0.0257^{+0.0140}_{-0.0108}$. The contour in
Fig.~\ref{fig:22combined} indicates a preference for both sets of
data. At the same time, the side located on the area of Planck
data allows for higher values of the tensor-to-scalar ratio, as
can be seen by pulling up. The contour shape is close to the oval
shape, unlike the similar case on the de Sitter background
(Fig.~\ref{fig:12combined}). At the same time, the shapes of the
contours are similar. This similarity indicates that the models
are identical, but thanks to the addition of $H_1$, the fitting
was carried out more efficiently. In comparison with the previous
power-law potential of the 2.1 model, it is noticeable that the
contour shapes are opposite, which creates prerequisites for
organic unification into a hybrid form.

\subsection{Hybrid Coupling}

\begin{figure}[ht]
\centering
\begin{minipage}{0.48\linewidth}
    \centering
    \includegraphics[width=\linewidth]{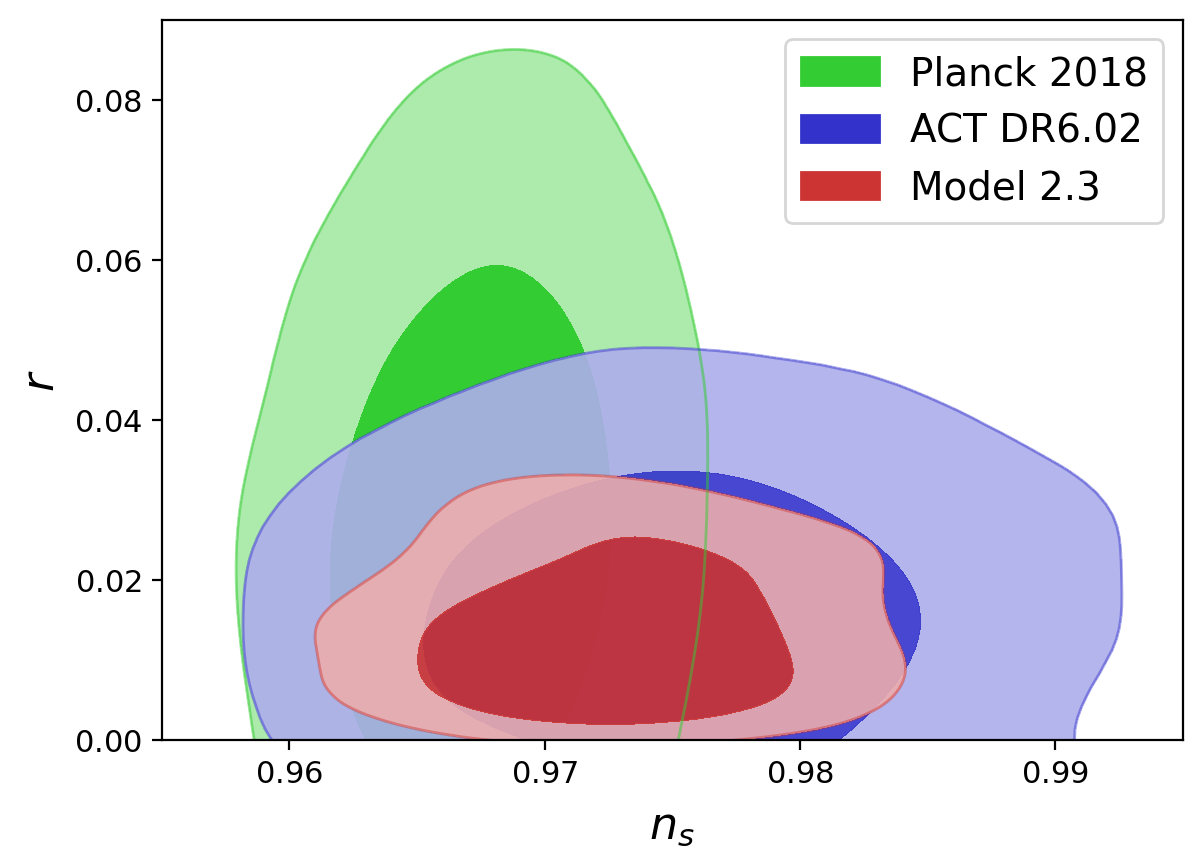}
\end{minipage}
\hfill
\begin{minipage}{0.48\linewidth}
    \centering
    \renewcommand{\arraystretch}{1.65}
    \begin{tabular}{cc}
    \toprule
    Parameter & Value \\
    \midrule
    $H_1$    & $0.0014^{+0.0011}_{-0.0010}$ \\
    $\mu$    & $0.1020^{+0.0051}_{-0.0065}$ \\
    $\gamma$ & $0.0060^{+0.0026}_{-0.0029}$ \\
    $b_1$    & $1.6064^{+0.2512}_{-0.3291}$ \\
    $b_2$    & $1.5387^{+0.3032}_{-0.3459}$ \\
    $n_s$    & $0.9727^{+0.0048}_{-0.0045}$ \\
    $r$      & $0.0118^{+0.0091}_{-0.0060}$ \\
    \bottomrule
    \end{tabular}
\end{minipage}
    \caption{Marginalized constraints for Model~2.3 on the $(n_s, r)$ plane
    at $N = 60$ for the hybrid coupling
    $h(\chi) = \gamma e^{b_1\chi}\chi^{b_2}$. The contours correspond to
    Planck~2018 (green) and ACT~DR6.02 (blue) at $1\sigma$ and $2\sigma$
    confidence levels. The table lists the best-fit parameter values
    with $1\sigma$ uncertainties.}
\label{fig:23constraints}
\end{figure}

The hybrid coupling function $h(\chi) =
\gamma\mathrm{e}^{b_1\chi}\chi^{b_2}$ on a quasi-de Sitter
background is a logical continuation of the analysis of the two
previous cases. As mentioned earlier, models 2.1 and 2.2 on the
$(n_s,r)$ plane are oriented opposite to each other. The contour
of the power-law model 2.1 tends to the lower left corner. The
contour of the exponential model 2.2, on the contrary, is
elongated to the upper right side. Such opposite orientations of
distribution suggests that combining the two functional forms
entails the unification and mutual compensation of asymmetries.
This is exactly the result demonstrated by
Fig.~\ref{fig:23constraints}.

\begin{figure}[ht]
\centering
\includegraphics[width=0.96\linewidth]{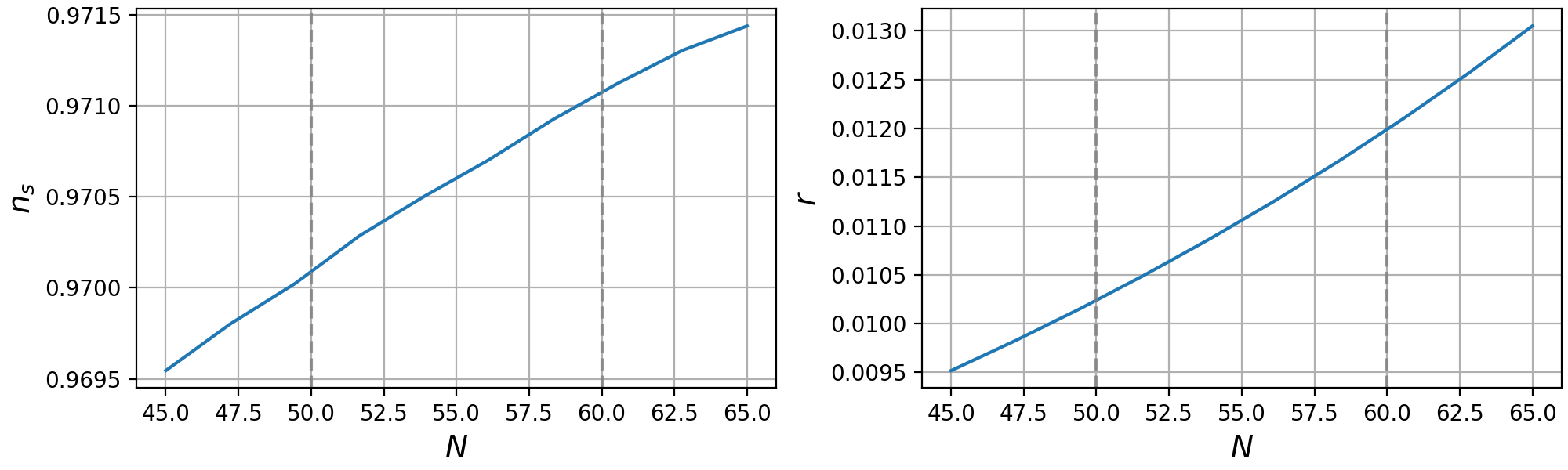}
\caption{Model~2.3 with hybrid coupling
$h(\chi) = \gamma e^{b_1\chi}\chi^{b_2}$ on a quasi-de~Sitter background
$H = H_0 - H_1 t$. The plots show the evolution of the spectral index $n_s(N)$
(left) and the tensor-to-scalar ratio $r(N)$ (right) over the inflationary
window $50 \leq N \leq 60$.}
\label{fig:23dynamics}
\end{figure}
The dynamics of $n_s(N)$ in the inflationary range of $50\leq
N\leq 60$ demonstrates monotonous growth, thereby arriving at a
more standard decreasing case. A similar evolution is observed for
the ratio $r(N)$. The values of the additional parameter $H_1 =
0.0014^{+0.0011}_{-0.0010}$ is average between the two previous
cases, which is consistent with the balancing nature of hybrid
coupling. The best parameter values for $N = 60$ are $n_s =
0.9727^{+0.0048}_{-0.0045}$ and $r = 0.0118^{+0.0091}_{-0.0060}$.
The parameter $\mu = 0.1020^{+0.0051}_{-0.0065}$ reproduces the
characteristic value of $\mu\approx 0.10$.

Balancing parameters $b_1 = 1.6064^{+0.2512}_{-0.3291}$ and $b_2 =
1.5387^{+0.3032}_{-0.3459}$ have approximately equal values. At
the same time, the errors of both parameters are lower compared to
the hybrid form of model 1.3, which indicates a better limitation
of the values. The contours in Fig.~\ref{fig:23constraints} are
close to the ideal oval shape, completely preferring the data of
ACT~DR6.02. Thus, the 2.3 model is the best in terms of selection
quality compared to other models of quasi-de Sitter background.

\subsection{Inverse Logarithmic Coupling}

\begin{figure}[ht]
    \centering
    \includegraphics[width=0.45\linewidth]{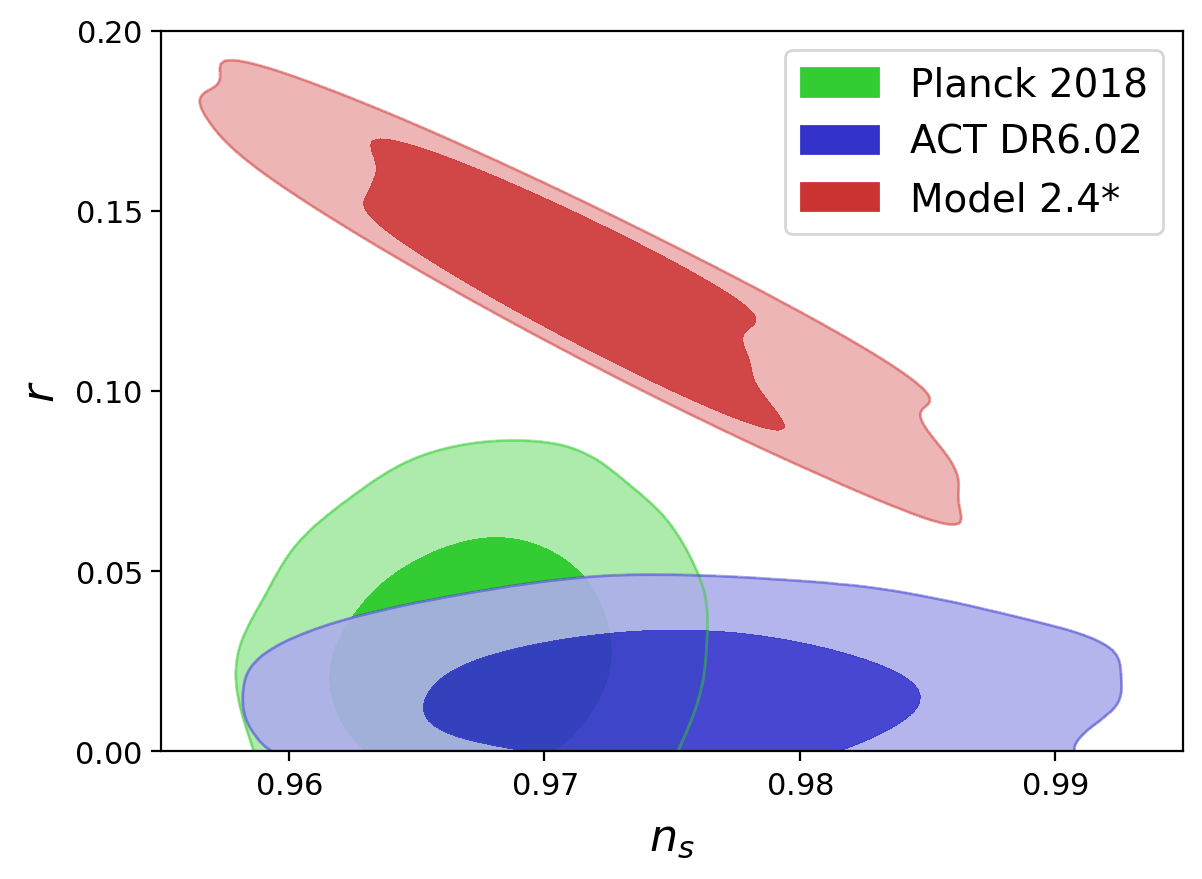}
    \caption{Marginalized constraints for Model~2.4* on the $(n_s, r)$ plane at
$N = 60$ for the direct logarithmic coupling $h(\chi) =
\gamma\ln(\chi/b)$. The contours correspond to Planck~2018 (green)
and ACT~DR6.02 (blue) at $1\sigma$ and $2\sigma$ confidence
levels.}
    \label{fig:240constraints}
\end{figure}

The direct logarithmic coupling $h(\chi) = \gamma\ln(\chi/b)$
against the de Sitter background (Fig.~\ref{fig:140nsr}) showed a
blue tilt of the scalar perturbation spectrum $n_s>1$. However the
additional parameter $H_1$ in quasi-de Sitter model can shift the
spectrum at certain values. To test this possibility, an MCMC
analysis was performed, the results of which are shown in
Fig.~\ref{fig:240constraints}

Even after significantly expanding the range of possible values,
the model does not match the observational data. The spectral
index of scalar perturbations has reached values
$n_s=0.9713^{+0.0064}_{-0.0059}$, which corresponds to both
datasets. However, the values of the tensor-to-scalar ratio
$r=0.1310^{+0.0228}_{-0.0239}$ are significantly higher than
observational. Thus, a direct logarithmic coupling is not viable
even against the background of quasi-de Sitter parameterization.

\begin{figure}[ht]
\centering
\includegraphics[width=0.96\linewidth]{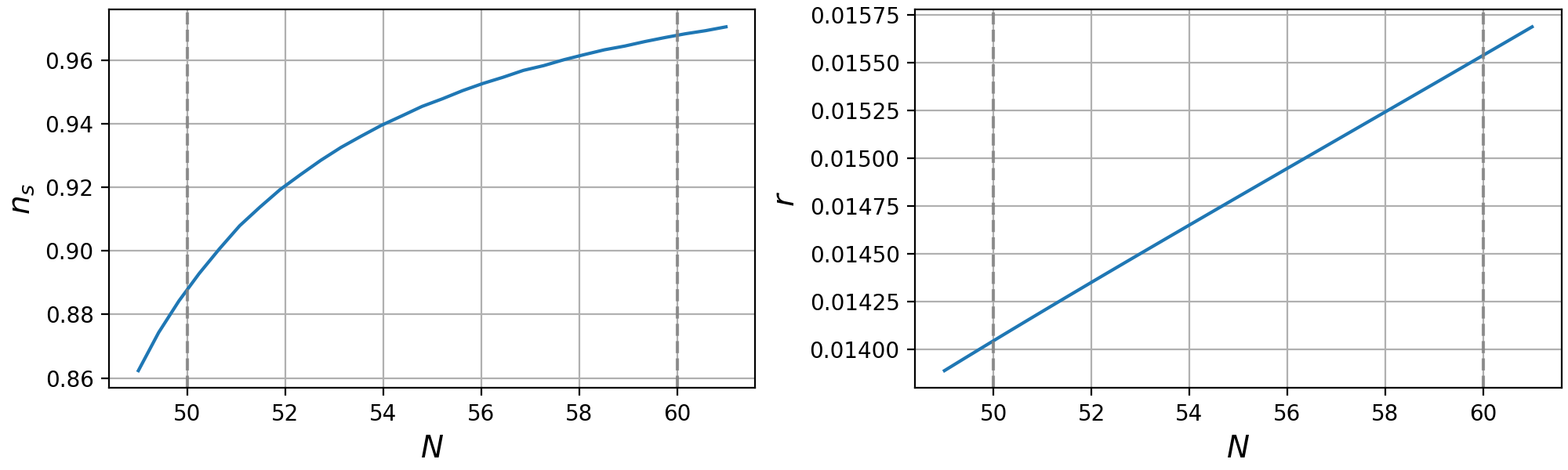}
\begin{minipage}{0.48\linewidth}
    \centering
    \includegraphics[width=\linewidth]{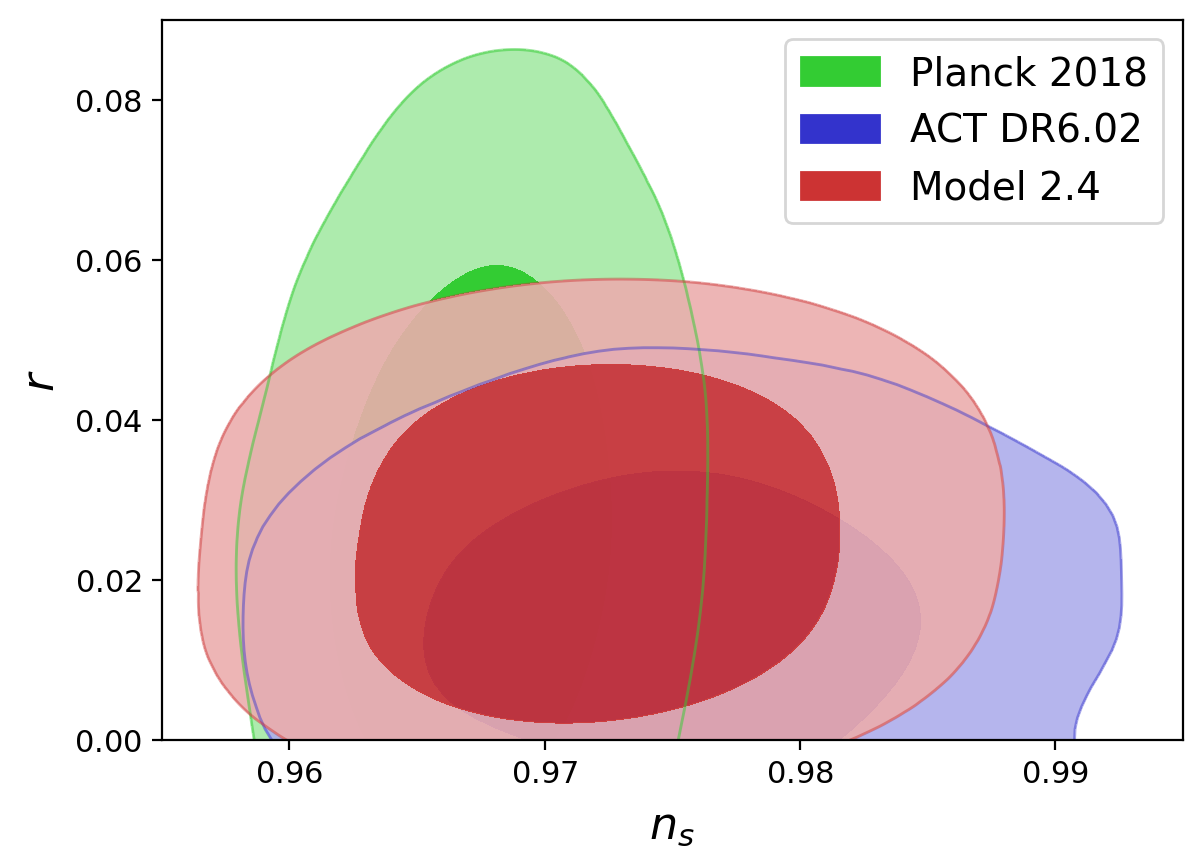}
\end{minipage}
\hfill
\begin{minipage}{0.48\linewidth}
    \centering
    \renewcommand{\arraystretch}{1.7}
    \begin{tabular}{cc}
    \toprule
    Parameter & Value \\
    \midrule
    $H_1$    & $0.0037^{+0.0009}_{-0.0014}$ \\
    $\mu$    & $0.1007^{+0.0060}_{-0.0068}$ \\
    $\gamma$ & $0.0048^{+0.0025}_{-0.0023}$ \\
    $b$      & $0.1689^{+0.0772}_{-0.0685}$ \\
    $n_s$    & $0.9726^{+0.0064}_{-0.0068}$ \\
    $r$      & $0.0262^{+0.0140}_{-0.0141}$ \\
    \bottomrule
    \space & \space
    \end{tabular}
\end{minipage}
\caption{Model~2.4: inverse logarithmic coupling
$h(\chi) = \gamma/\ln(\chi/b)$ on the quasi-de~Sitter background
$H = H_0 - H_1 t$. The dynamic plots show the evolution of
$n_s(N)$ (left) and $r(N)$ (right) over the inflationary window
$50 \leq N \leq 60$. The contour plot shows marginalized
constraints on the $(n_s,\, r)$ plane at $N = 60$ against
Planck~2018 (green) and ACT~DR6.02 (blue) at $1\sigma$ and
$2\sigma$ confidence levels. The table lists best-fit parameter
values $\pm1\sigma$.}
\label{fig:24combined}
\end{figure}

For this reason, the inverse logarithmic relationship $h(\chi) =
\gamma/\ln(\chi/b)$ is being investigated. The value of the
parameter $H_1 = 0.0037^{+0.0009}_{-0.0014}$ significantly more
than the previous cases, which indicates a greater deviation from
the de Sitter regime. The best parameter values for $N = 60$ are
$n_s = 0.9726^{+0.0064}_{-0.0068}$ and $r =
0.0262^{+0.0140}_{-0.0141}$. The contour of the model in
Fig.~\ref{fig:24combined} indicates a preference for both
datasets, while having a uniform oval distribution, indicating a
high quality of parameter selection.

\section{Results: Exponential Hubble Evolution}\label{sec:exp}

\subsection{Power-Law Coupling}

\begin{figure}[htb]
\centering
\includegraphics[width=0.96\linewidth]{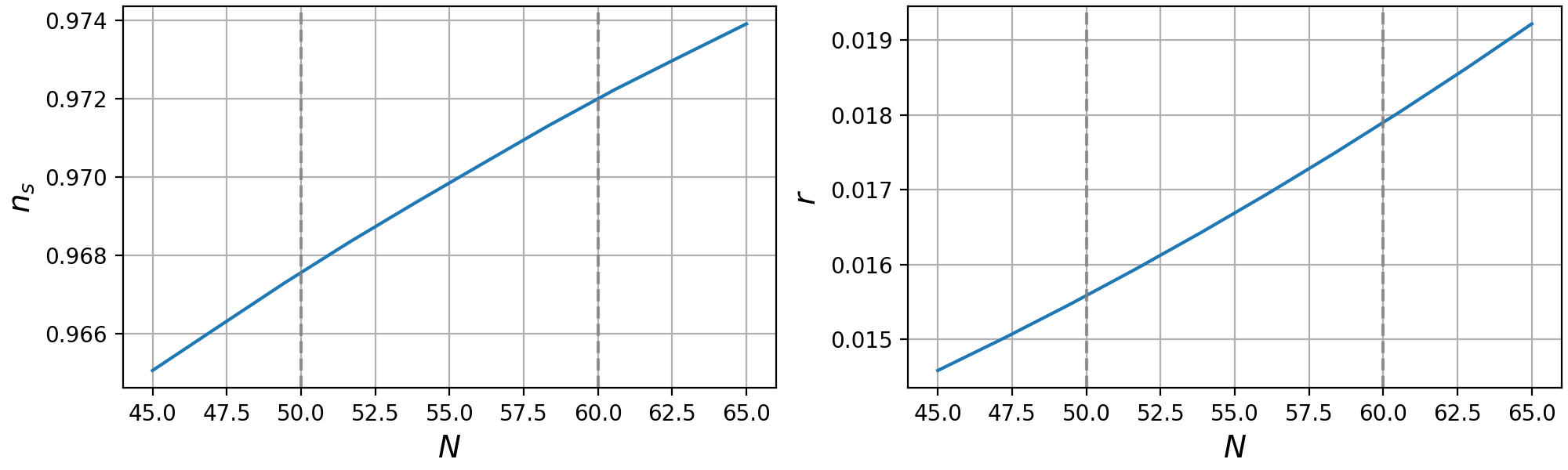}
\caption{Model~3.1: power-law coupling $h(\chi) = (\chi/M)^n$
on the exponential background $H = H_0\mathrm{e}^{-\Omega t}$. The plots show the evolution of the spectral index $n_s(N)$ (left) and the tensor-to-scalar ratio $r(N)$ (right) over the inflationary window $50 \leq N \leq 60$.}
\label{fig:31dynamics}
\end{figure}

The power-law coupling function $h(\chi) =(\chi/M)^n$ on an
exponential background $H = H_0 e^{-\Omega t}$, unlike previous
similar cases (Model 1.1, 2.1), uses the normalization scale $M$
instead of the parameter $\gamma$. The parameterization of the
Hubble function contains the parameter $\Omega$, which controls
the degree of deviation from the de Sitter case. The calculation
results indicate small, but not zero values of the parameter
$\Omega = 0.0025^{+0.0018}_{-0.0015}$. Fig.~\ref{fig:31dynamics}
demonstrates a monotonous growth of the functions of the spectral
index of scalar perturbations $n_s(N)$ and the tensor-to-scalar
ratio $r(N)$.

\begin{figure}[htb]
\centering
\begin{minipage}{0.48\linewidth}
    \centering
    \includegraphics[width=\linewidth]{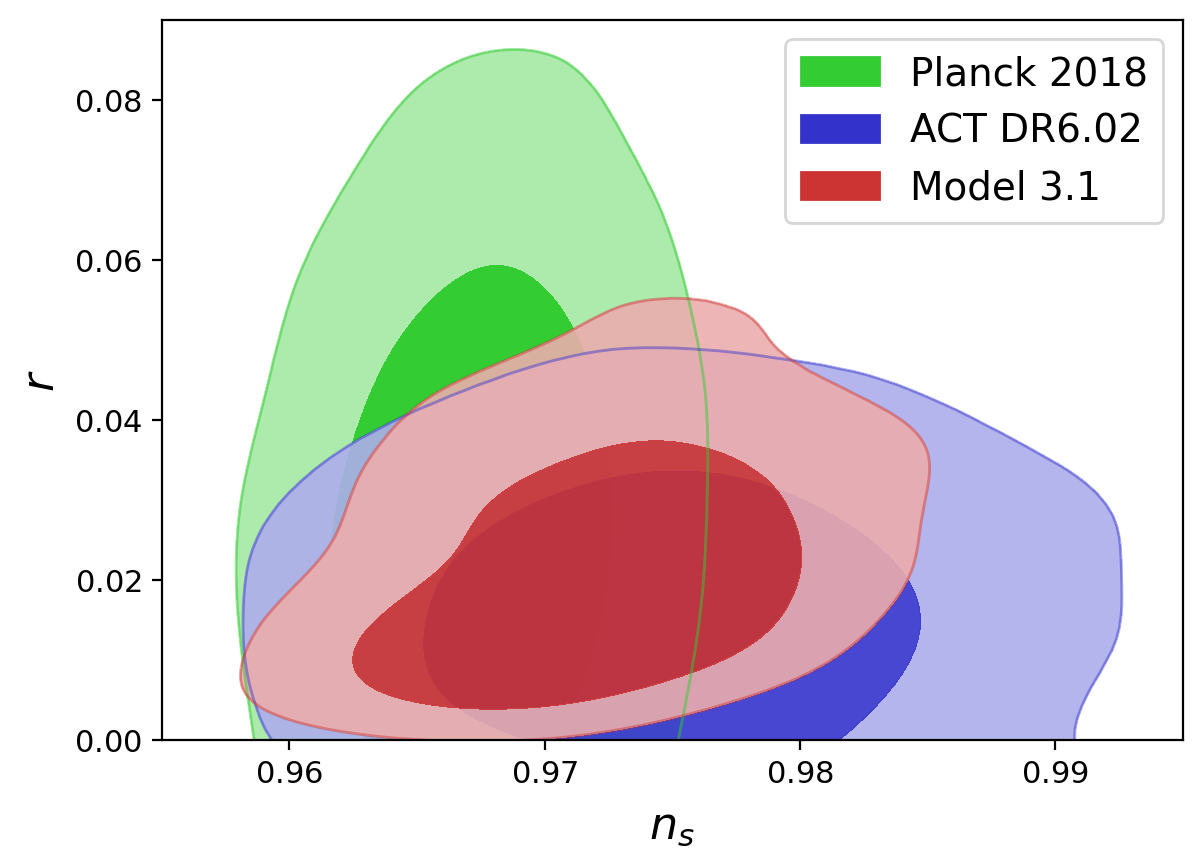}
\end{minipage}
\hfill
\begin{minipage}{0.48\linewidth}
    \centering
    \renewcommand{\arraystretch}{1.7}
    \begin{tabular}{cc}
    \toprule
    Parameter & Value \\
    \midrule
    $\Omega$ & $0.0025^{+0.0018}_{-0.0015}$ \\
    $\mu$    & $0.1006^{+0.0060}_{-0.0063}$ \\
    $M$      & $2.9368^{+0.5988}_{-0.5792}$ \\
    $n$      & $3.2067^{+0.4418}_{-0.4431}$ \\
    $n_s$    & $0.9720^{+0.0053}_{-0.0054}$ \\
    $r$      & $0.0189^{+0.0134}_{-0.0091}$ \\
    \bottomrule
    \end{tabular}
\end{minipage}
\caption{Model~3.1: power-law coupling $h(\chi) = (\chi/M)^n$
on the exponential background $H = H_0\mathrm{e}^{-\Omega t}$.
The contours correspond to Planck~2018 (green) and ACT~DR6.02 (blue) at $1\sigma$ and $2\sigma$ confidence levels. The table lists the best-fit parameter values with $1\sigma$ uncertainties.}
\label{fig:31constraints}
\end{figure}

The value of the scale parameter $M = 2.9368^{+0.5988}_{-0.5792}$
and the power parameter $n = 3.2067^{+0.4418}_{-0.4431}$ indicate
a moderately rapid growth of the coupling function in the
inflationary range of e-folds. The contour plot in
Fig.~\ref{fig:31constraints} demonstrates the model's preference
for both datasets. Stretching the contour in the lower left corner
shows similarities with previous models of the power-law type. The
best parameter values at $N = 60$ are $n_s =
0.9720^{+0.0053}_{-0.0054}$ and $r = 0.0189^{+0.0134}_{-0.0091}$.

\subsection{Exponential Coupling}

\begin{figure}[htb]
\centering
\includegraphics[width=0.96\linewidth]{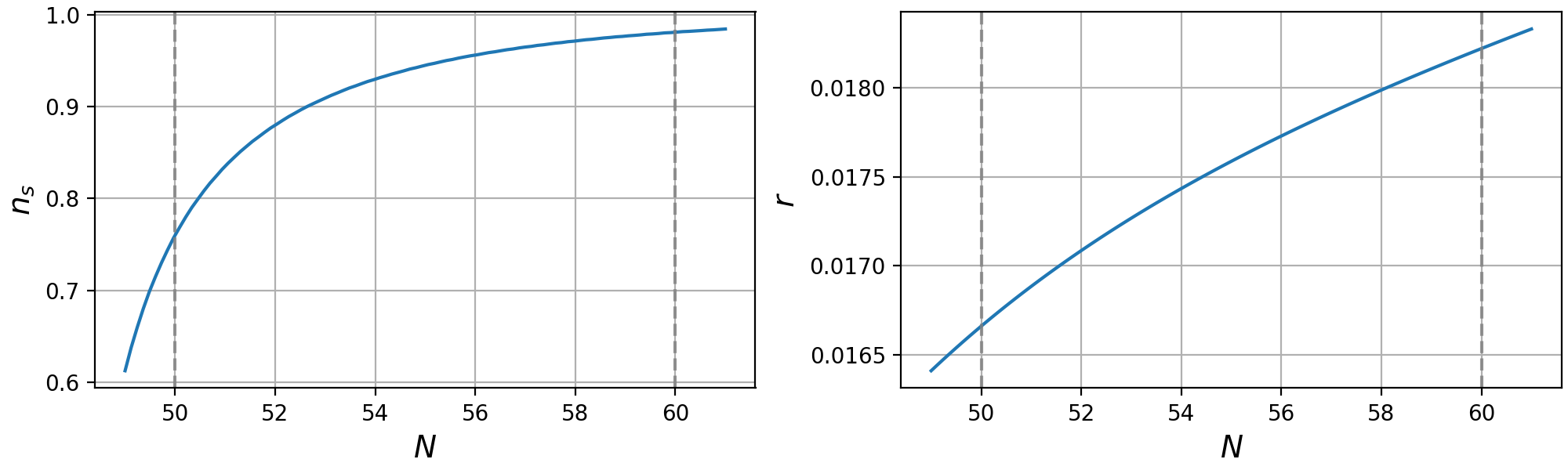}
\begin{minipage}{0.48\linewidth}
    \centering
    \includegraphics[width=\linewidth]{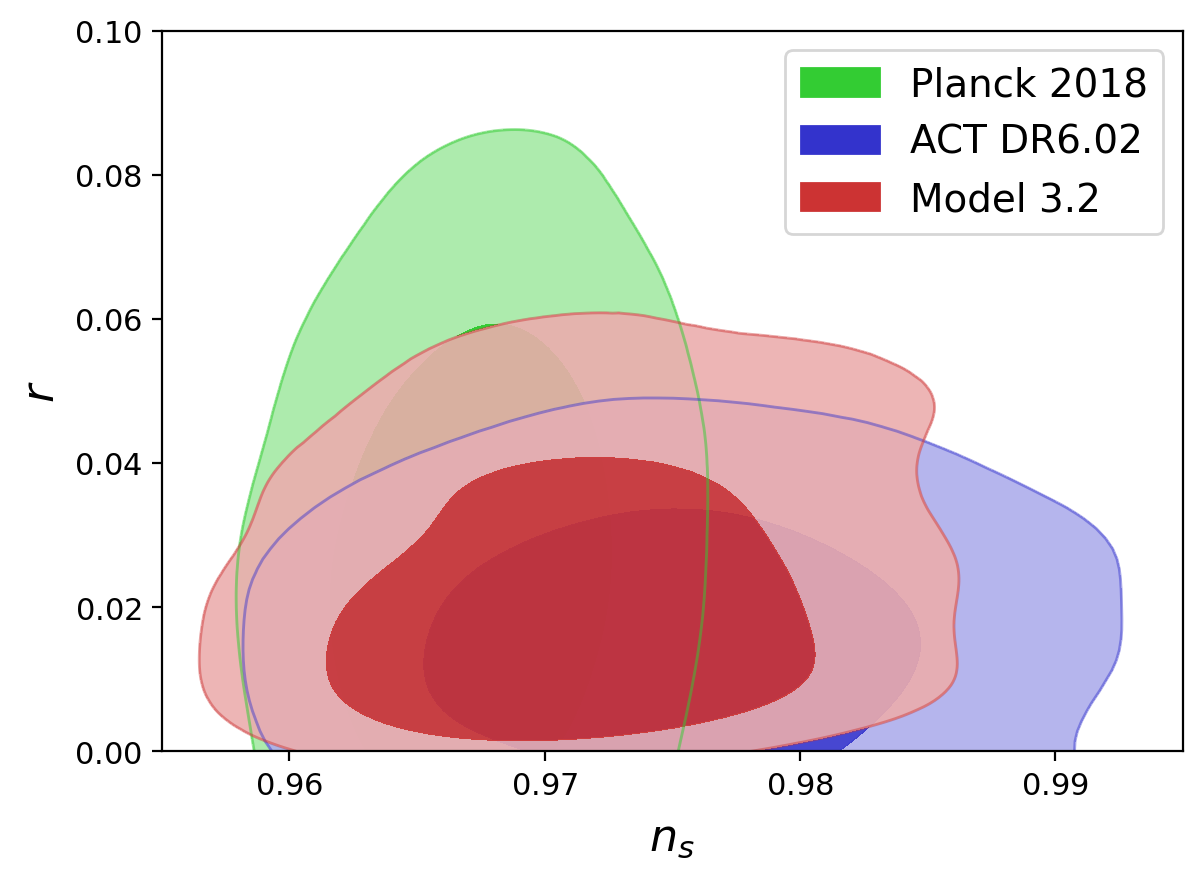}
\end{minipage}
\hfill
\begin{minipage}{0.48\linewidth}
    \centering
    \renewcommand{\arraystretch}{1.7}
    \begin{tabular}{cc}
    \toprule
    Parameter & Value \\
    \midrule
    $\Omega$ & $0.0038^{+0.0021}_{-0.0018}$ \\
    $\mu$    & $0.0986^{+0.0068}_{-0.0056}$ \\
    $\alpha$ & $13.4674^{+2.8266}_{-2.1996}$ \\
    $n_s$    & $0.9720^{+0.0063}_{-0.0067}$ \\
    $r$      & $0.0195^{+0.0168}_{-0.0106}$ \\
    \bottomrule
    \space & \space
    \end{tabular}
\end{minipage}
\caption{Model~3.2: exponential coupling
$h(\chi) = \mathrm{e}^{-\alpha\chi}$ on the exponential
background $H = H_0\mathrm{e}^{-\Omega t}$. The dynamic plots
show the evolution of $n_s(N)$ (left) and $r(N)$ (right) over
the inflationary window $50 \leq N \leq 60$. The contour plot
shows marginalized constraints on the $(n_s,\, r)$ plane at
$N = 60$ against Planck~2018 (green) and ACT~DR6.02 (blue) at
$1\sigma$ and $2\sigma$ confidence levels. The table lists
best-fit parameter values $\pm1\sigma$.}
\label{fig:32combined}
\end{figure}

The exponential coupling function $h(\chi)= e^{-\alpha\chi}$
against an exponential background $H = H_0 e^{-\Omega t}$
demonstrates a significant increase in the spectral index of
scalar perturbations $n_s(N)$. Its values increases in the range
$[0.75, 0.97]$ (Fig.~\ref{fig:32combined}), which is a
non-standard case of dynamics. The tensor-to-scalar ratio, on the
contrary, demonstrates monotonous growth. Unlike the previous
cases, the power of $-\alpha\chi$ is negative. This choice is due
to the linear growth of the function $\chi=\mu^2 t$, which in the
exponential case of the coupling function leads to an uncontrolled
increase in the contribution of the Gauss-Bonnet invariant and the
blue tilt of the scalar perturbation spectrum. A negative degree
allows for setting the decreasing dynamics of the coupling
function, which weakens the influence of the Gauss-Bonnet
invariant. Despite the rapid growth of the spectral index, the
values correspond to the observational data, preferring both
datasets.

The best parameter values for $N = 60$ are $n_s =
0.9720^{+0.0063}_{-0.0067}$ and $r = 0.0195^{+0.0168}_{-0.0106}$.
The contour in Fig.~\ref{fig:32combined} is close to an oval shape
and has wide light areas, which indicates a good selection of
parameters. Large values of the parameter $\alpha =
13.4674^{+2.8266}_{-2.1996}$ indicate a rapid suppression of the
contribution of the Gauss-Bonnet invariant at the end of
inflation. The value of $\Omega = 0.0038^{+0.0021}_{-0.0018}$
comparatively more than in model 3.1, which leads to a greater
deviation from the de Sitter case.

\subsection{Hybrid Coupling}

\begin{figure}[ht]
\centering
\includegraphics[width=0.96\linewidth, keepaspectratio]{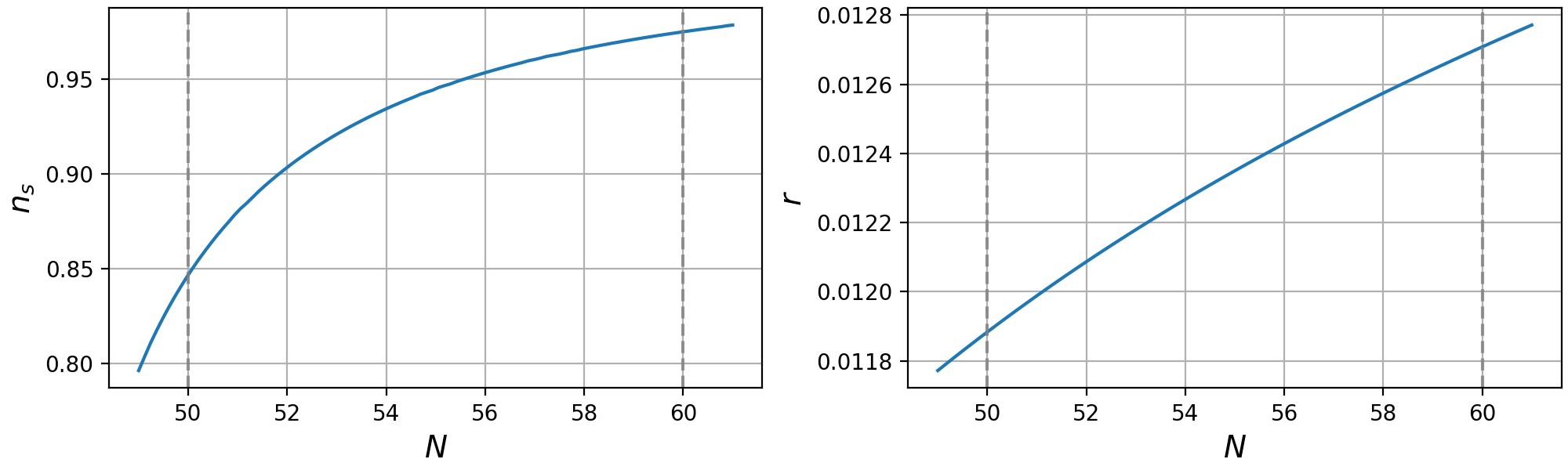}
\begin{minipage}{0.48\linewidth}
    \centering
    \includegraphics[width=\linewidth]{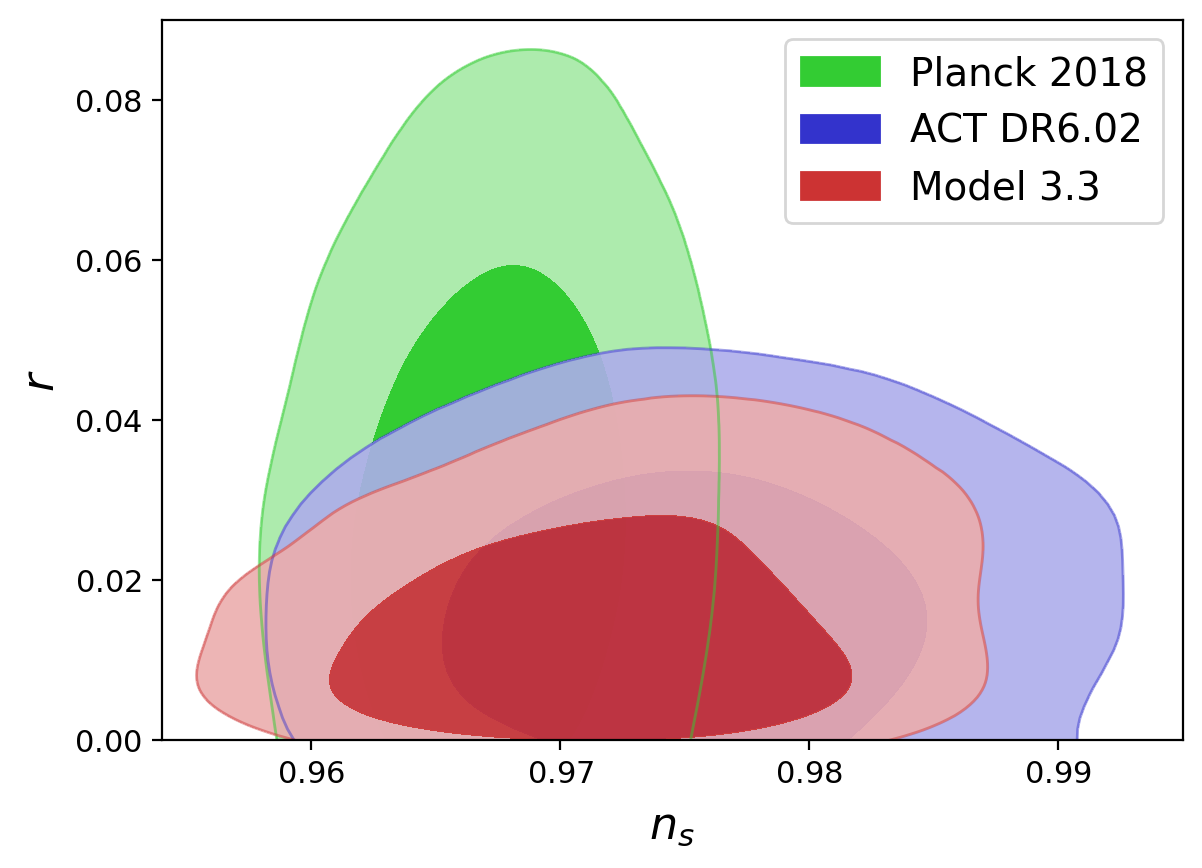}
\end{minipage}
\hfill
\begin{minipage}{0.48\linewidth}
    \centering
    \renewcommand{\arraystretch}{1.7}
    \begin{tabular}{cc}
    \toprule
    Parameter & Value \\
    \midrule
    $\Omega$ & $0.0027^{+0.0018}_{-0.0014}$ \\
    $\mu$    & $0.0983^{+0.0062}_{-0.0056}$ \\
    $\alpha$ & $12.9981^{+2.6594}_{-2.0102}$ \\
    $n$      & $1.4711^{+0.3136}_{-0.3066}$ \\
    $n_s$    & $0.9724^{+0.0065}_{-0.0070}$ \\
    $r$      & $0.0127^{+0.0115}_{-0.0071}$ \\
    \bottomrule
    \space & \space
    \end{tabular}
\end{minipage}
\caption{Model~3.3: hybrid coupling
$h(\chi) = \mathrm{e}^{-\alpha\chi}\chi^n$ on the exponential
background $H = H_0\mathrm{e}^{-\Omega t}$. The dynamic plots
show the evolution of $n_s(N)$ (left) and $r(N)$ (right) over
the inflationary window $50 \leq N \leq 60$. The contour plot
shows marginalized constraints on the $(n_s,\, r)$ plane at
$N = 60$ against Planck~2018 (green) and ACT~DR6.02 (blue) at
$1\sigma$ and $2\sigma$ confidence levels. The table lists
best-fit parameter values $\pm1\sigma$.}
\label{fig:33combined}
\end{figure}

Hybrid coupling function $h(\chi) = e^{-\alpha\chi}\chi^n$ on
exponential background $H= H_0 e^{-\Omega t}$ is designed to
balance the excessive growth rate of the exponential coupling
function $h(\chi) = e^{-\alpha\chi}$. The power-law coupling
$h(\chi) =(\chi/M)^n$ slows down the growth rate of the spectral
index of scalar perturbations, and the exponential coupling
ensures a rapid decrease in the influence of Gauss-Bonnet
invariants at the end of inflation. The effectiveness of this
approach is demonstrated in Fig.~\ref{fig:33combined}, where the
spectral index increases in the range [0.85, 0.97]. The
tensor-to-scalar ratio demonstrates monotonous growth.

The values of $\alpha = 12.9981^{+2.6594}_{-2.0102}$ are close to
the analogous value in the model 3.2, indicating the dominant
contribution of the exponential part. Values of the parameter
$\Omega = 0.0027^{+0.0018}_{-0.0014}$ are small, which indicates a
small deviation from the de Sitter expansion. The best parameter
values for $N = 60$ are $n_s = 0.9724^{+0.0065}_{-0.0070}$ and $r
= 0.0127^{+0.0115}_{-0.0071}$. The contour plot in
Fig.~\ref{fig:33combined} demonstrates good parameter selection
quality, as can be seen from the oval contour with wide light
areas. The model has the same preference for both datasets, while
the quality of the selection is better compared to the exponential
case.

\subsection{Inverse Logarithmic Coupling}

\begin{figure}[htb]
    \centering
    \includegraphics[width=0.96\linewidth]{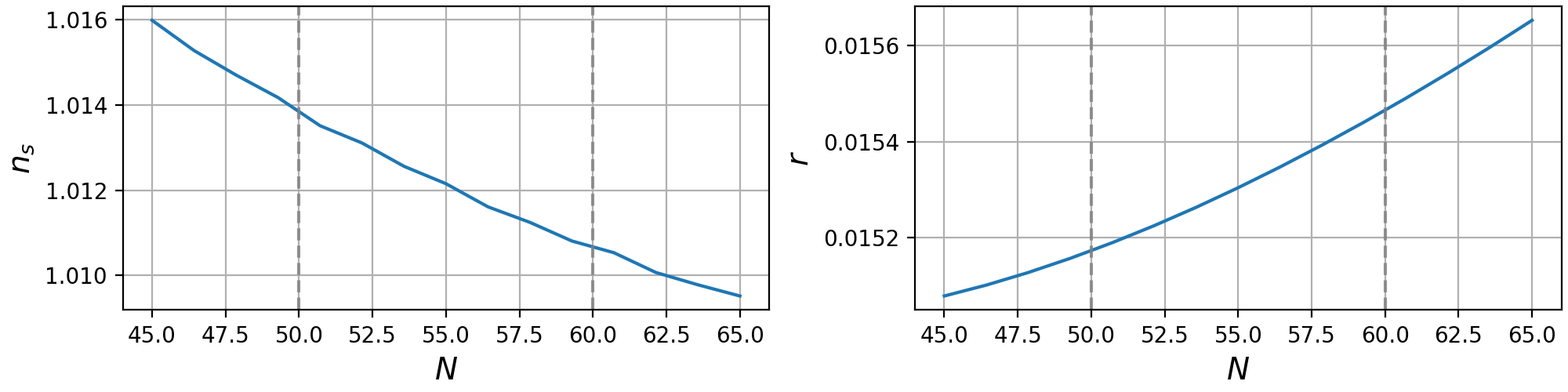}
    \caption{The dynamics of the spectral index of scalar perturbations $n_s(N)$
    and the tensor-to-scalar ratio $r(N)$ in the inflationary range
    $50\leq N\leq 60$ with a direct logarithmic coupling
    $h(\chi) = \gamma\ln(\chi/b)$.}
    \label{fig:340nsr}
\end{figure}

\begin{figure}[ht]
\centering
\includegraphics[width=0.96\linewidth]{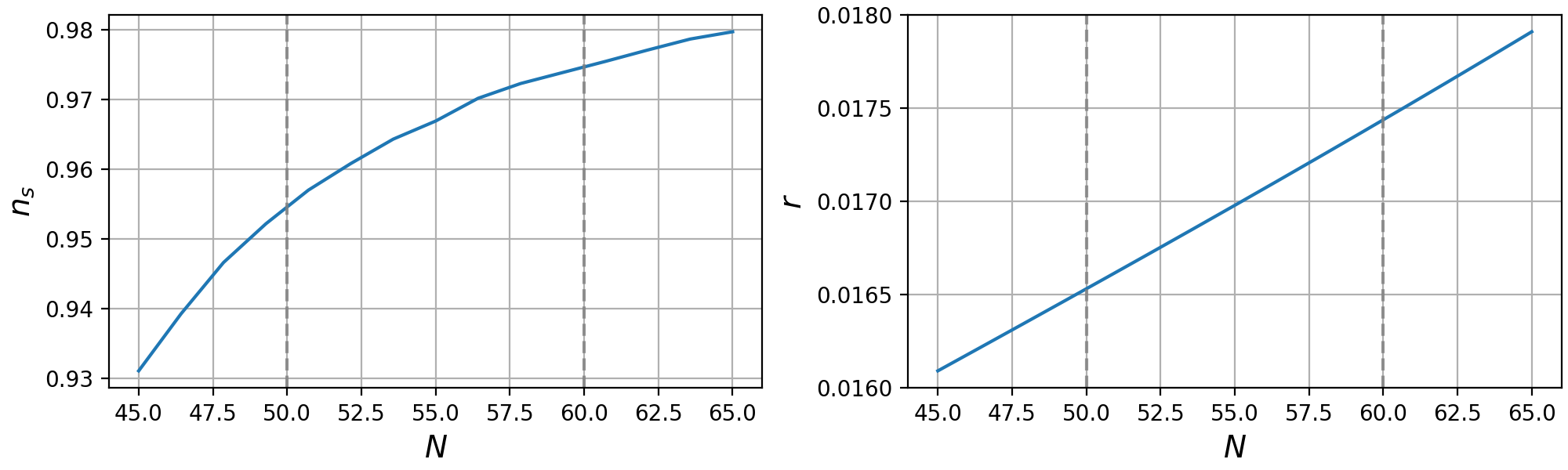}
\begin{minipage}{0.48\linewidth}
    \centering
    \includegraphics[width=\linewidth]{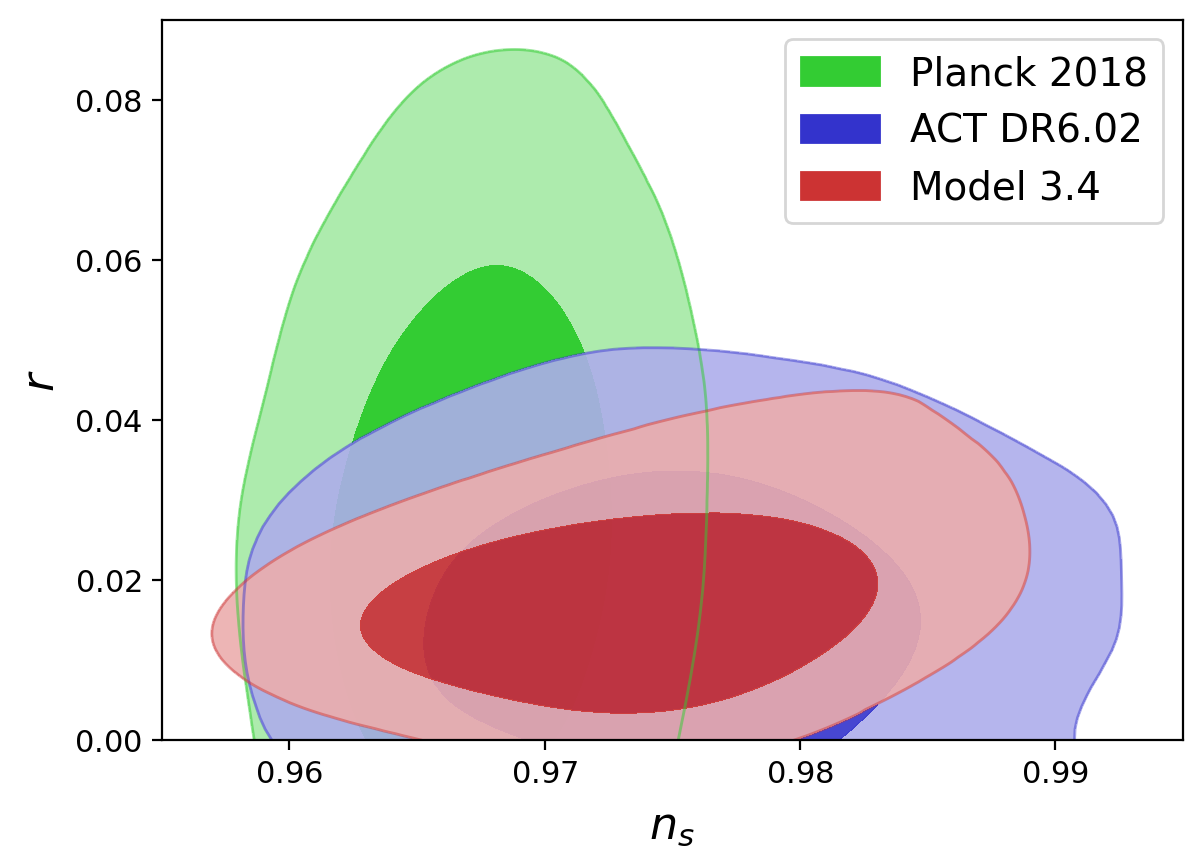}
\end{minipage}
\hfill
\begin{minipage}{0.48\linewidth}
    \centering
    \renewcommand{\arraystretch}{1.7}
    \begin{tabular}{cc}
    \toprule
    Parameter & Value \\
    \midrule
    $\Omega$ & $0.0036^{+0.0011}_{-0.0011}$ \\
    $\mu$    & $0.1005^{+0.0062}_{-0.0063}$ \\
    $\gamma$ & $0.0351^{+0.0095}_{-0.0129}$ \\
    $b$      & $10.0483^{+3.1092}_{-3.2797}$ \\
    $n_s$    & $0.9746^{+0.0064}_{-0.0072}$ \\
    $r$      & $0.0176^{+0.0074}_{-0.0063}$ \\
    \bottomrule
    \space & \space
    \end{tabular}
\end{minipage}
\caption{Model~3.4: inverse logarithmic coupling
$h(\chi) = \gamma/\ln(\chi/b)$ on the exponential background
$H = H_0\mathrm{e}^{-\Omega t}$. The dynamic plots show the
evolution of $n_s(N)$ (left) and $r(N)$ (right) over the
inflationary window $50 \leq N \leq 60$. The contour plot shows
marginalized constraints on the $(n_s,\, r)$ plane at $N = 60$
against Planck~2018 (green) and ACT~DR6.02 (blue) at $1\sigma$
and $2\sigma$ confidence levels. The table lists best-fit
parameter values $\pm1\sigma$.}
\label{fig:34combined}
\end{figure}

As in the previous cases, the direct logarithmic coupling $h(\chi)
= \gamma\ln(\chi/b)$ on an exponential background $H = H_0
e^{-\Omega t}$ for any parameter values demonstrates the blue tilt
of the scalar perturbation spectrum $n_s > 1$, which is clearly
demonstrated in Fig.~\ref{fig:340nsr}. At the same time, the
tensor-to-scalar ratio demonstrates values corresponding to the
observational data. For this reason, the inverse logarithmic
coupling $h(\chi) = \gamma/\ln(\chi/b)$ is investigated, the
calculation results of which are shown in
Fig.~\ref{fig:34combined}. The best parameter values at $N = 60$
are $n_s = 0.9746^{+0.0064}_{-0.0072}$ and $r =
0.0176^{+0.0074}_{-0.0063}$.

The contour plot in Fig.~\ref{fig:34combined} demonstrates the
viability of the model and the good quality of parameter
selection. The model prefers both datasets in about the same way.
Values of the $\Omega = 0.0036^{+0.0011}_{-0.0011}$ indicate a
slight deviation from the constant expansion of de Sitter case.
The parameter $\mu = 0.1005^{+0.0062}_{-0.0063}$ reproduces the
characteristic value $\mu\approx 0.10$, consistently observed in
all model configurations.

\section{Results: Fractional Hubble Parameter}\label{sec:frac}

The fractional form of the Hubble parameter $H =n/t$ is inspired
by the classical scenario of power-law inflation, which is usually
given by the scale factor $a(t)\propto t^n$. This parametrization
occupies a special place in the research of inflation, as it
provides a significantly faster decrease over time, unlike in the
cases of de Sitter and quasi-de Sitter background. The rate of
decrease is set by the parameter $n \gg 1$. An important feature
of this parametrization is the natural lower bound of the
tensor-to-scalar ratio, which is observed in most of the following
configurations. This effect is due to the structure of the
slow-roll parameters. For example, the first parameter
$\epsilon_1=- \dot{H}/H^2 =  1/n$ takes fixed values that depend
only on the parameter $n$. Despite the fact that fractional
parametrization was not investigated in the original work, its
inclusion in the inflationary analysis is necessary for a
full-fledged study.

\subsection{Power-Law Coupling}

The power-law coupling function on a fractional background $H =
n/t$ is given as $h(\chi) = \gamma\chi^b$. The most noticeable
feature of this configuration is the large values of $n =
309.2433^{+121.4744}_{-123.9947}$ with wide confidence intervals.
It follows from this that the model is close to exponential
expansion, but at the same time retains fundamental mathematical
differences. Wide confidence intervals indicate low sensitivity to
such large parameter values.

\begin{figure}[ht]
\centering
\includegraphics[width=0.96\linewidth]{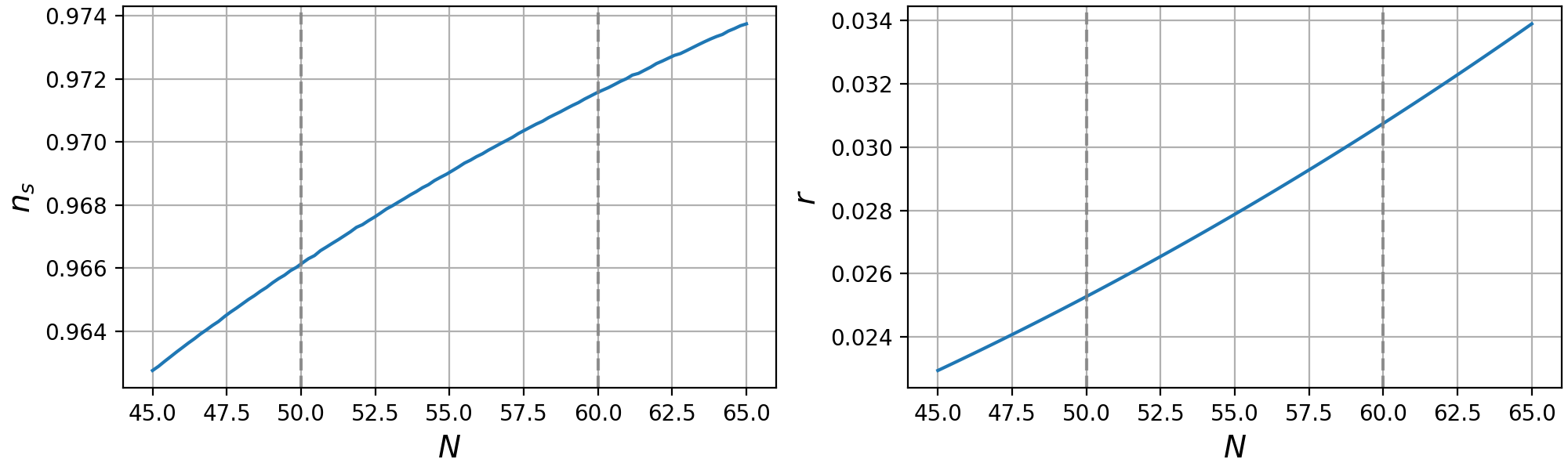}
\begin{minipage}{0.48\linewidth}
    \centering
    \includegraphics[width=\linewidth]{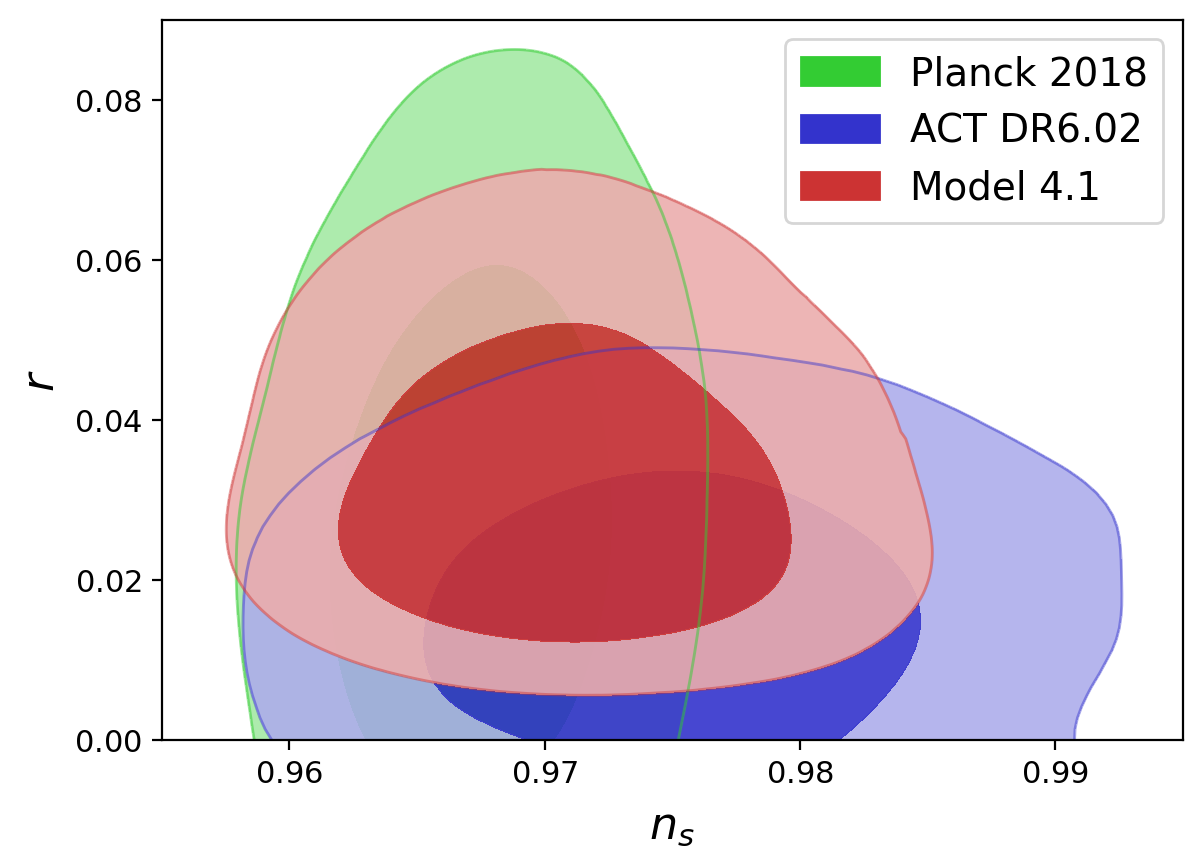}
\end{minipage}
\hfill
\begin{minipage}{0.48\linewidth}
    \centering
    \renewcommand{\arraystretch}{1.7}
    \begin{tabular}{cc}
    \toprule
    Parameter & Value \\
    \midrule
    $\mu$    & $0.1003^{+0.0062}_{-0.0066}$ \\
    $n$      & $309.2433^{+121.4744}_{-123.9947}$ \\
    $\gamma$ & $0.0929^{+0.0477}_{-0.0488}$ \\
    $b$      & $3.1595^{+0.3357}_{-0.3708}$ \\
    $n_s$    & $0.9706^{+0.0056}_{-0.0056}$ \\
    $r$      & $0.0299^{+0.0152}_{-0.0105}$ \\
    \bottomrule
    \space & \space
    \end{tabular}
\end{minipage}
\caption{Model~4.1: power-law coupling $h(\chi) = \gamma\chi^b$
on the fractional background $H = n/t$. The dynamic plots show
the evolution of $n_s(N)$ (left) and $r(N)$ (right) over the
inflationary window $50 \leq N \leq 60$. The contour plot shows
marginalized constraints on the $(n_s,\, r)$ plane at $N = 60$
against Planck~2018 (green) and ACT~DR6.02 (blue) at $1\sigma$
and $2\sigma$ confidence levels. The table lists best-fit
parameter values $\pm1\sigma$.}
\label{fig:41combined}
\end{figure}

The dynamics of the spectral index of scalar perturbations
$n_s(N)$ demonstrates monotonous growth, as does the
tensor-to-scalar ratio. The best parameter values for $N = 60$ are
$n_s = 0.9706^{+0.0056}_{-0.0056}$ and $r =
0.0299^{+0.0152}_{-0.0105}$. It can be seen that the values of the
tensor-to-scalar ratio are significantly higher than in most
previous cases and exceed the upper limit of the ACT DR6.02 data.
Despite the large values, the upper limit of the Planck 2018 data
has not been exceeded. The contour plot in
Fig.~\ref{fig:41combined} has an oval shape with wide light areas.
Despite the wide confidence intervals of the $n$ parameter, the
model demonstrates good selection quality and prefers the Planck
2018 data. It can be seen that the area 2$\sigma$ is narrower from
below than from the other sides, which indicates the presence of a
limitation of the tensor-to-scalar ratio. The parameter $\mu =
0.1003^{+0.0062}_{-0.0066}$ reproduces the characteristic value of
$\mu\approx 0.10$, which is observed in all configurations of the
model.

\subsection{Exponential Coupling}

\begin{figure}[ht]
\centering
\includegraphics[width=0.96\linewidth]{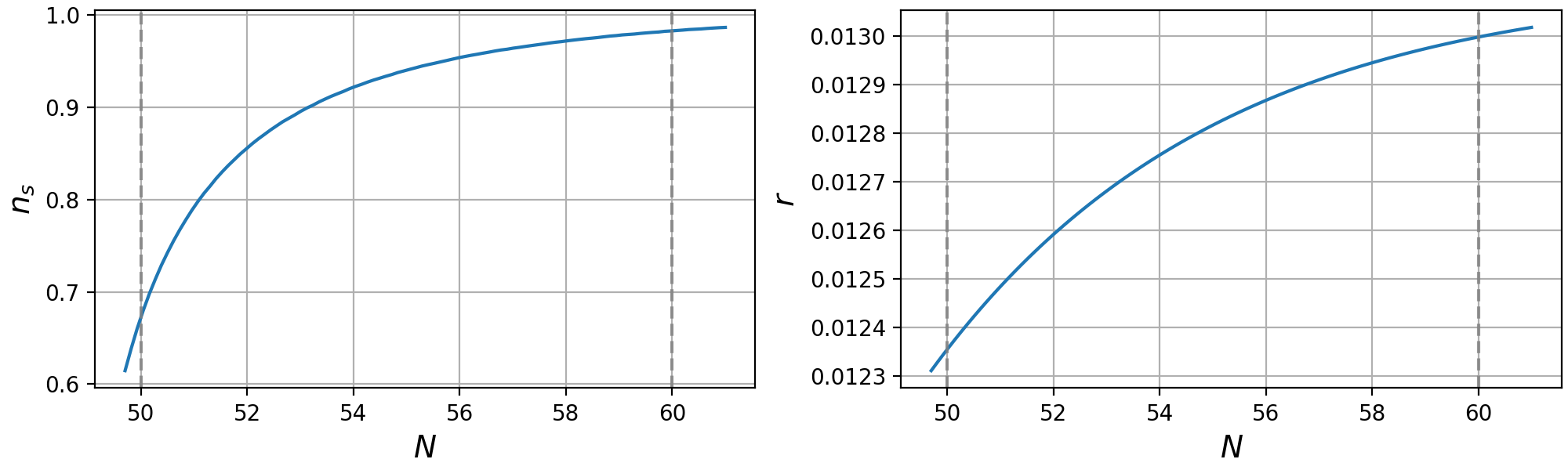}
\begin{minipage}{0.48\linewidth}
    \centering
    \includegraphics[width=\linewidth]{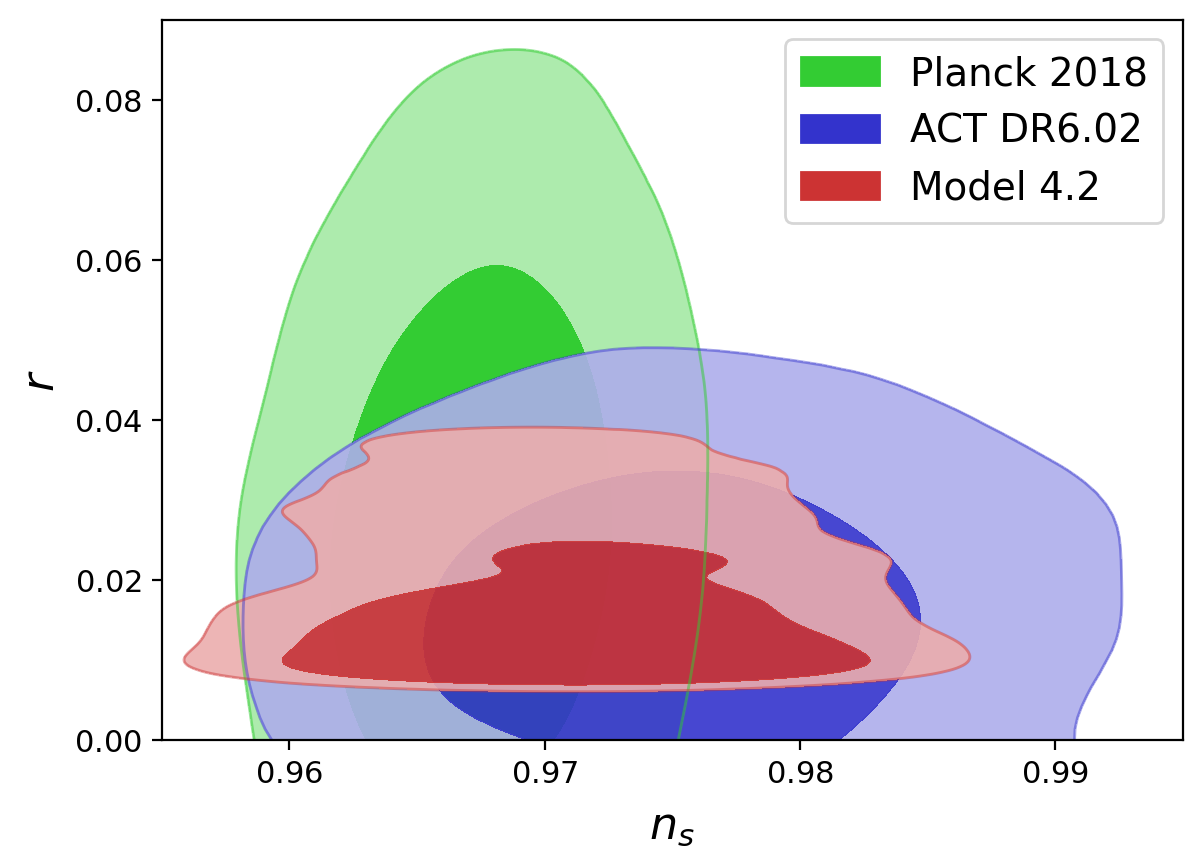}
\end{minipage}
\hfill
\begin{minipage}{0.48\linewidth}
    \centering
    \renewcommand{\arraystretch}{1.7}
    \begin{tabular}{cc}
    \toprule
    Parameter & Value \\
    \midrule
    $\mu$    & $0.1023^{+0.0048}_{-0.0055}$ \\
    $\gamma$ & $1.4575^{+0.3504}_{-0.2891}$ \\
    $n$      & $305.0783^{+121.1494}_{-115.9408}$ \\
    $b$      & $14.2113^{+1.5780}_{-1.7437}$ \\
    $n_s$    & $0.9713^{+0.0063}_{-0.0063}$ \\
    $r$      & $0.0129^{+0.0080}_{-0.0037}$ \\
    \bottomrule
    \space & \space
    \end{tabular}
\end{minipage}
\caption{Model~4.2: exponential coupling
$h(\chi) = \gamma\mathrm{e}^{-b\chi}$ on the fractional
background $H = n/t$. The dynamic plots show the evolution of
$n_s(N)$ (left) and $r(N)$ (right) over the inflationary window
$50 \leq N \leq 60$. The contour plot shows marginalized
constraints on the $(n_s,\, r)$ plane at $N = 60$ against
Planck~2018 (green) and ACT~DR6.02 (blue) at $1\sigma$ and
$2\sigma$ confidence levels. The table lists best-fit parameter
values $\pm1\sigma$.}
\label{fig:42combined}
\end{figure}

Exponential coupling function $h(\chi) = \gamma e^{-b\chi}$ on a
fractional background $H = n/t$ has several notable features.
Similarly to the exponential coupling from Model 3.2, a negative
degree of $-b\chi$ is used here, which prevents an uncontrolled
increase in the contribution of the Gauss-Bonnet invariant. As in
Model 3.2, an extremely rapid increase in the spectral index of
scalar perturbations $n_s(N)$ is observed, as can be seen in
Fig.~\ref{fig:42combined}. The tensor-to-scalar ratio $r(N)$, on
the contrary, is extremely weakly dependent on the number of
e-folds. The contour plot also differs from the previous cases.
The contour has an uneven distribution, as can be seen from the
ragged edges. Despite the poor quality of the parameter selection,
the model is still in the range of values of both datasets, which
indicates its viability. The characteristic asymptotics of the
lower bound of the tensor-to-scalar ratio can also be traced in
this case. The best parameter values at $N = 60$ are $n_s =
0.9713^{+0.0063}_{-0.0063}$ and $r = 0.0129^{+0.0080}_{-0.0037}$.

\subsection{Hybrid Coupling}

\begin{figure}[htb]
\centering
\includegraphics[width=0.96\linewidth]{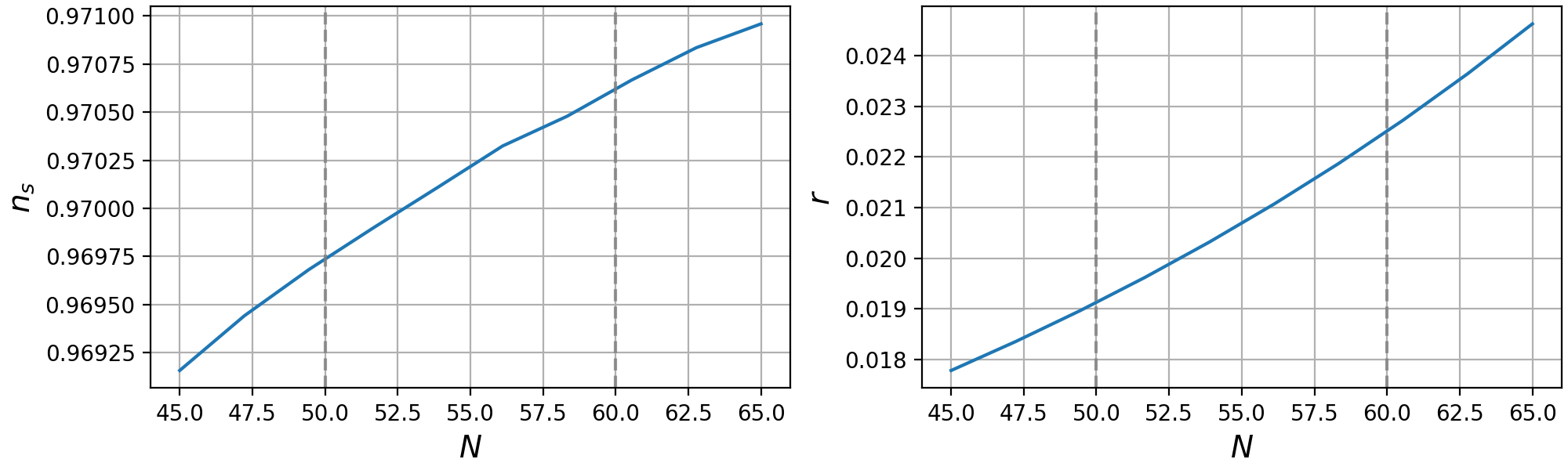}
\caption{Model~4.3: hybrid coupling $h(\chi) = \gamma e^{-b_1\chi}\chi^{b_2}$
on the fractional background $H = n/t$. The plots show the evolution of the
spectral index $n_s(N)$ (left) and the tensor-to-scalar ratio $r(N)$ (right)
over the inflationary window $50 \leq N \leq 60$.}
\label{fig:43dynamics}
\end{figure}

\begin{figure}[htb]
\centering
\begin{minipage}{0.48\linewidth}
    \centering
    \includegraphics[width=\linewidth]{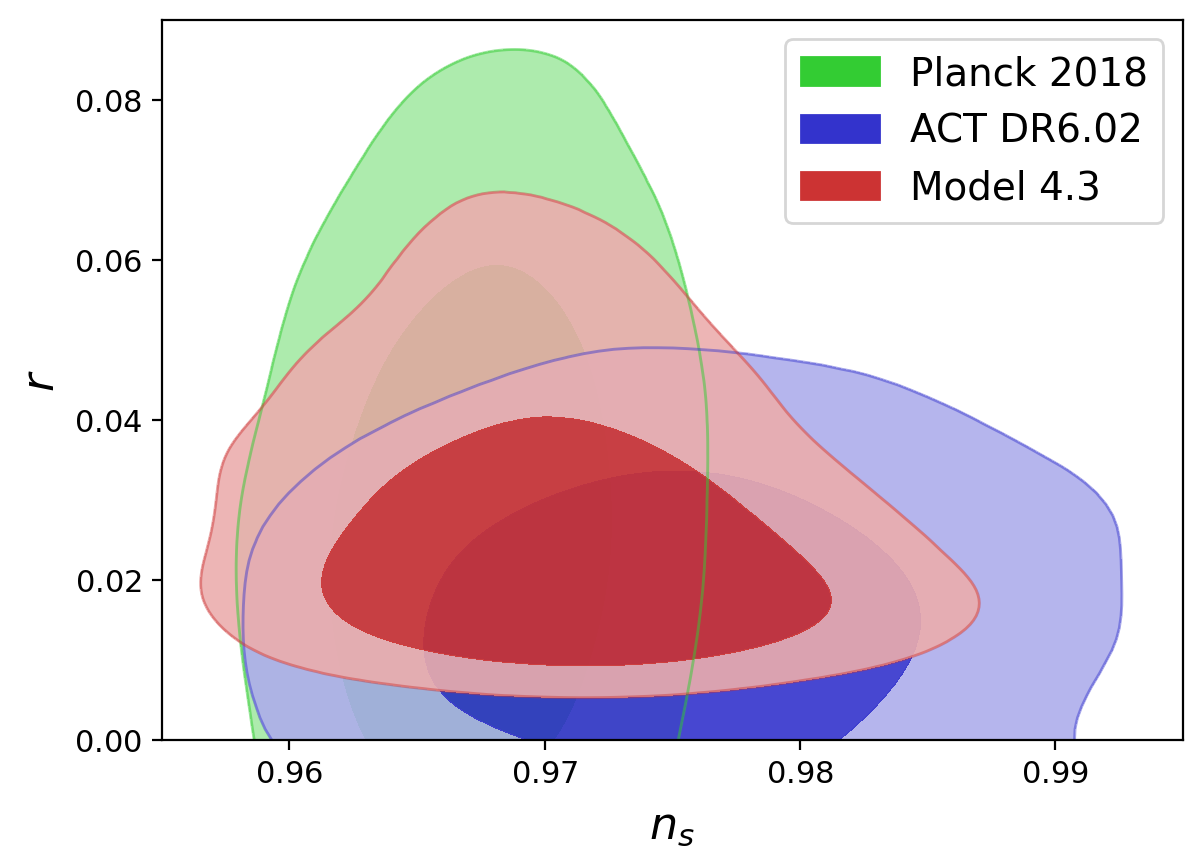}
\end{minipage}
\hfill
\begin{minipage}{0.48\linewidth}
    \centering
    \renewcommand{\arraystretch}{1.7}
    \begin{tabular}{cc}
    \toprule
    Parameter & Value \\
    \midrule
    $\mu$    & $0.1000^{+0.0061}_{-0.0062}$ \\
    $n$      & $354.5086^{+94.7474}_{-92.8283}$ \\
    $\gamma$ & $0.0149^{+0.0077}_{-0.0080}$ \\
    $b_1$    & $1.699^{+0.6319}_{-0.7619}$ \\
    $b_2$    & $1.661^{+0.8063}_{-0.8303}$ \\
    $n_s$    & $0.9707^{+0.0056}_{-0.0058}$ \\
    $r$      & $0.0226^{+0.0143}_{-0.0075}$ \\
    \bottomrule
    \end{tabular}
\end{minipage}
\caption{Model~4.3: hybrid coupling $h(\chi) = \gamma e^{
b_1\chi}\chi^{b_2}$ on the fractional background $H = n/t$. The
contours correspond to Planck~2018 (green) and ACT~DR6.02 (blue)
at $1\sigma$ and $2\sigma$ confidence levels. The table lists the
best-fit parameter values with $1\sigma$ uncertainties.}
\label{fig:43constraints}
\end{figure}

The hybrid coupling function $h(\chi) = \gamma
e^{-b_1\chi}\chi^{b_2}$ on a fractional background $H = n/t$ is
designed to reduce the sharp increase in the spectral index of
scalar perturbations and smooth the contour, providing a better
selection of parameters. The effectiveness of this approach is
evident in Fig.~\ref{fig:43dynamics}, where the growth rate of the
spectral index $n_s(N)$ has noticeably decreased compared to the
previous case. The hybrid form also made it possible to increase
the quality of parameter selection, as evidenced by the contour in
Fig.~\ref{fig:43constraints}, which has smoother edges. The hybrid
model prefers the Planck 2018 data, as evidenced by the large
values of the tensor-to-scalar ratio and the overall shift towards
a redder tilt of the spectrum. Balancing parameters of $b_1 =
1.699^{+0.6319}_{-0.7619}$ and $b_2 = 1.661^{+0.8063}_{-0.8303}$
have similar values, which indicates the equal contribution of the
exponential and power parts. The best parameter values at $N = 60$
are $n_s = 0.9707^{+0.0056}_{-0.0058}$ and $r =
0.0226^{+0.0143}_{-0.0075}$.

\subsection{Inverse Logarithmic Coupling}

\begin{figure}[htb]
\centering
\includegraphics[width=0.96\linewidth]{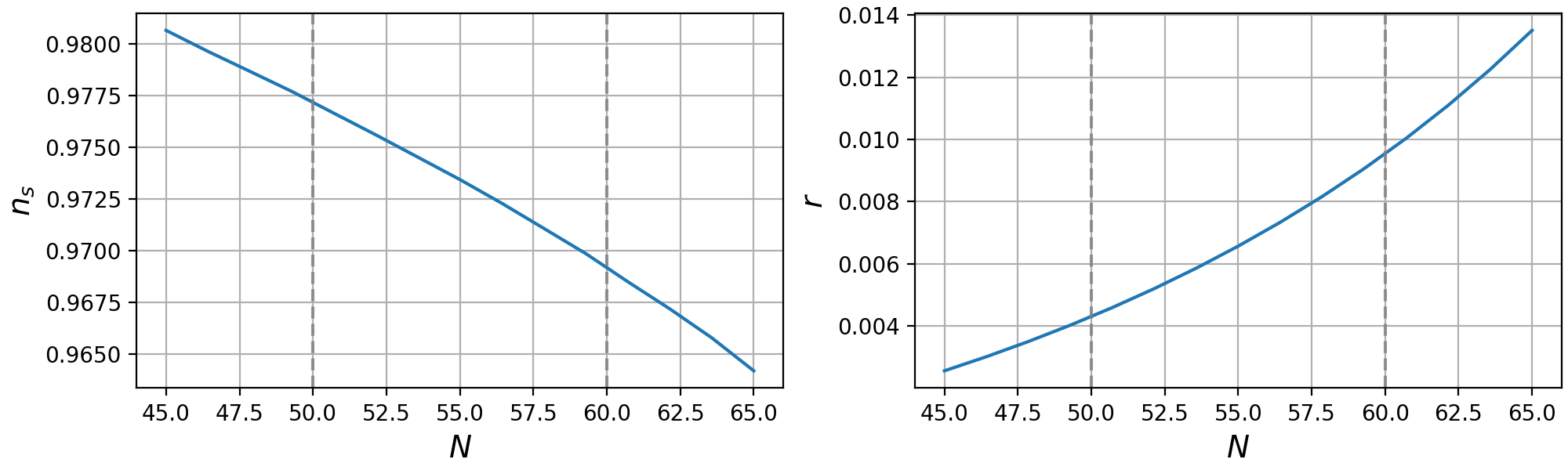}
\begin{minipage}{0.48\linewidth}
    \centering
    \includegraphics[width=\linewidth]{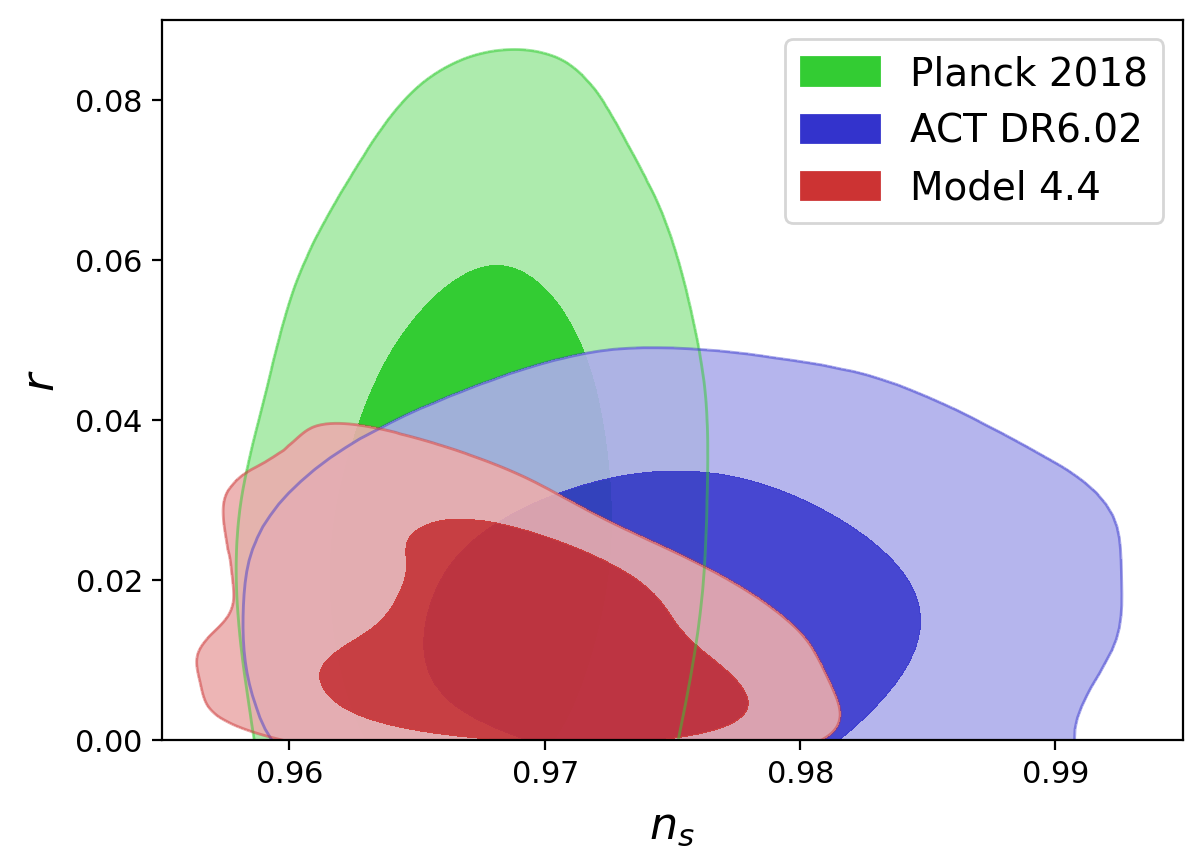}
\end{minipage}
\hfill
\begin{minipage}{0.48\linewidth}
    \centering
    \renewcommand{\arraystretch}{1.7}
    \begin{tabular}{cc}
    \toprule
    Parameter & Value \\
    \midrule
    $\mu$    & $0.0965^{+0.0047}_{-0.0040}$ \\
    $\gamma$ & $0.0546^{+0.0296}_{-0.0280}$ \\
    $n$      & $309.9986^{+56.3473}_{-64.5982}$ \\
    $b$      & $1.2616^{+0.0946}_{-0.1104}$ \\
    $n_s$    & $0.9689^{+0.0053}_{-0.0053}$ \\
    $r$      & $0.0092^{+0.0122}_{-0.0064}$ \\
    \bottomrule
    \space & \space
    \end{tabular}
\end{minipage}
\caption{Model~4.4: inverse logarithmic coupling
$h(\chi) = \gamma/\ln(\chi/b)$ on the fractional background
$H = n/t$. The dynamic plots show the evolution of $n_s(N)$
(left) and $r(N)$ (right) over the inflationary window
$50 \leq N \leq 60$. The contour plot shows marginalized
constraints on the $(n_s,\, r)$ plane at $N = 60$ against
Planck~2018 (green) and ACT~DR6.02 (blue) at $1\sigma$ and
$2\sigma$ confidence levels. The table lists best-fit parameter
values $\pm1\sigma$.}
\label{fig:44combined}
\end{figure}

The direct logarithmic coupling $h(\chi) = \gamma\ln(\chi/b)$, as
in all previous cases, demonstrates the blue tilt of the spectrum
of scalar perturbations, which definitively excludes it from the
list of viable forms within the framework of ghost-free $f(R,
\mathcal{G})$ model. The inverse logarithmic coupling $h(\chi) =
\gamma/\ln(\chi/b)$ provides the red tilt of the spectrum ($n_s =
0.9689^{+0.0053}_{-0.0053}$). However, unlike other models with a
fractional background $H = n/t$, the tensor-to-scalar ratio does
not have an asymptotic lower bound ($r =
0.0092^{+0.0122}_{-0.0064}$). The contour plot in
Fig.~\ref{fig:44combined} demonstrates the acceptable quality of
parameter selection and the viability of the model.

\section{Summary and Conclusions}\label{sec:discussion}

\begin{figure}[ht]
    \centering
    \begin{tabular}{cc}
        \includegraphics[width=0.48\linewidth]{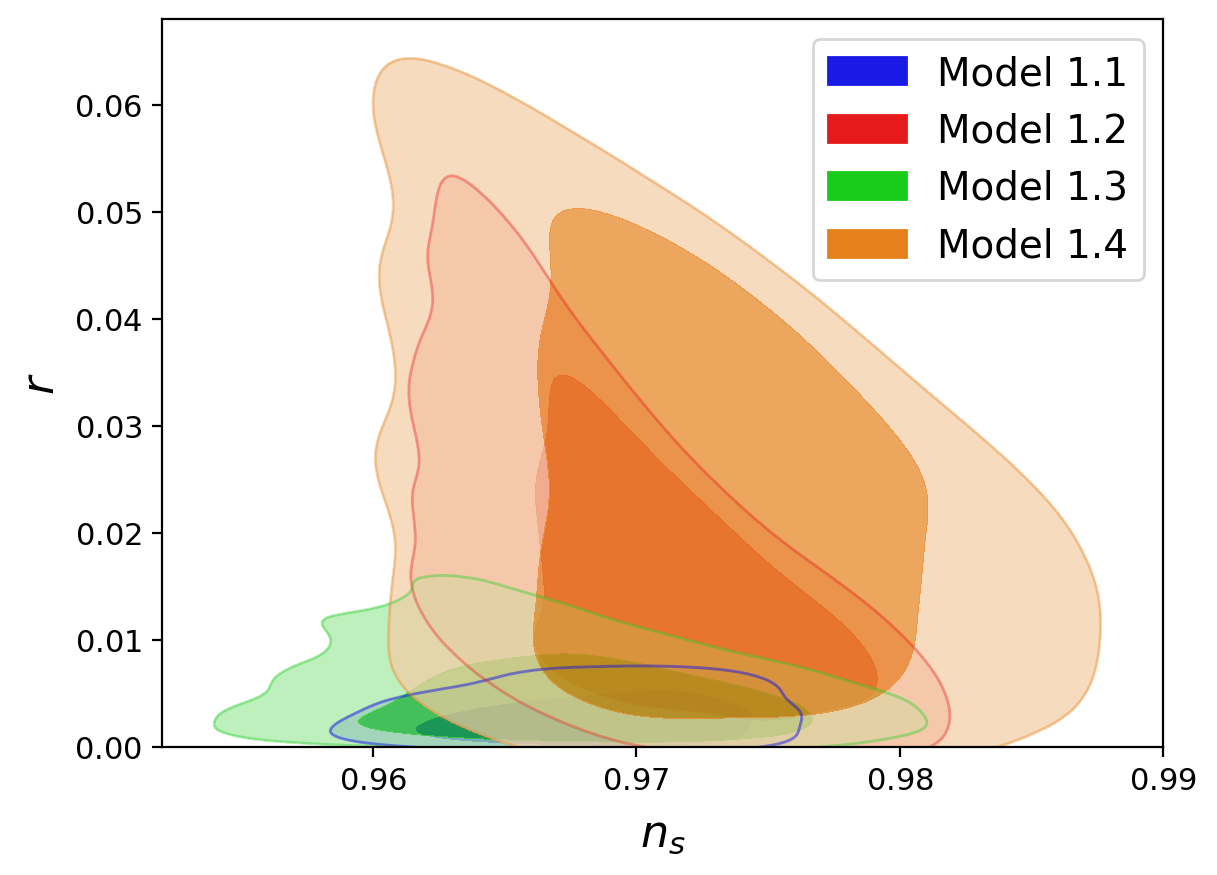} &
        \includegraphics[width=0.48\linewidth]{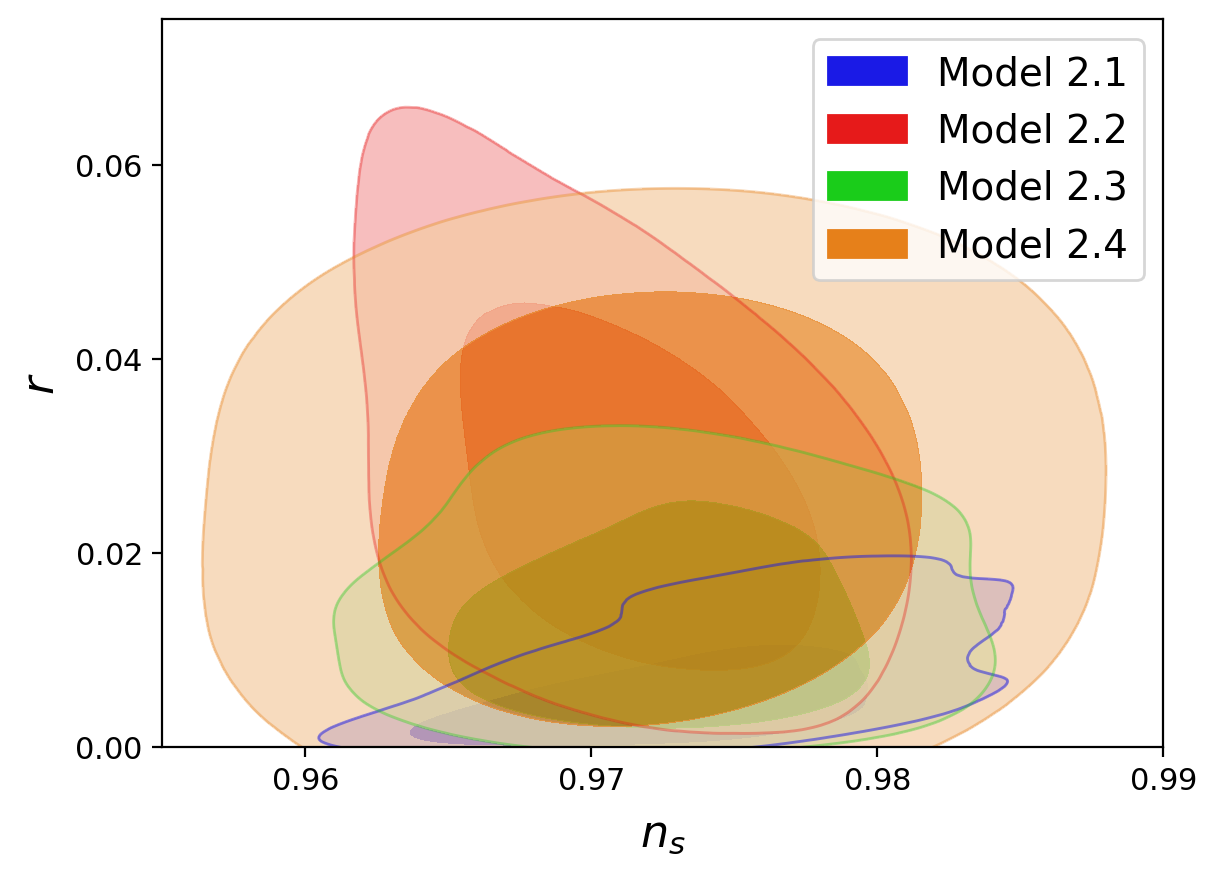} \\
        (a) & (b) \\[6pt]
        \includegraphics[width=0.48\linewidth]{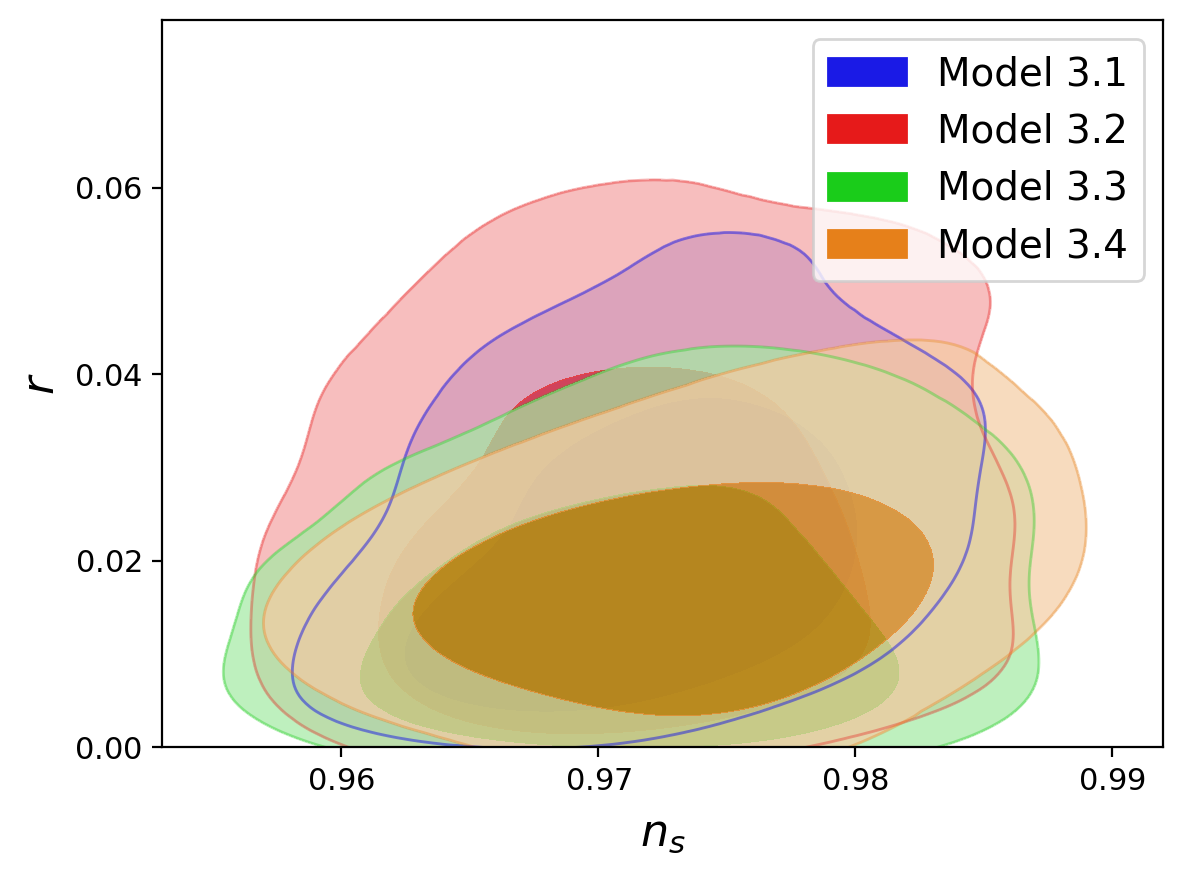} &
        \includegraphics[width=0.48\linewidth]{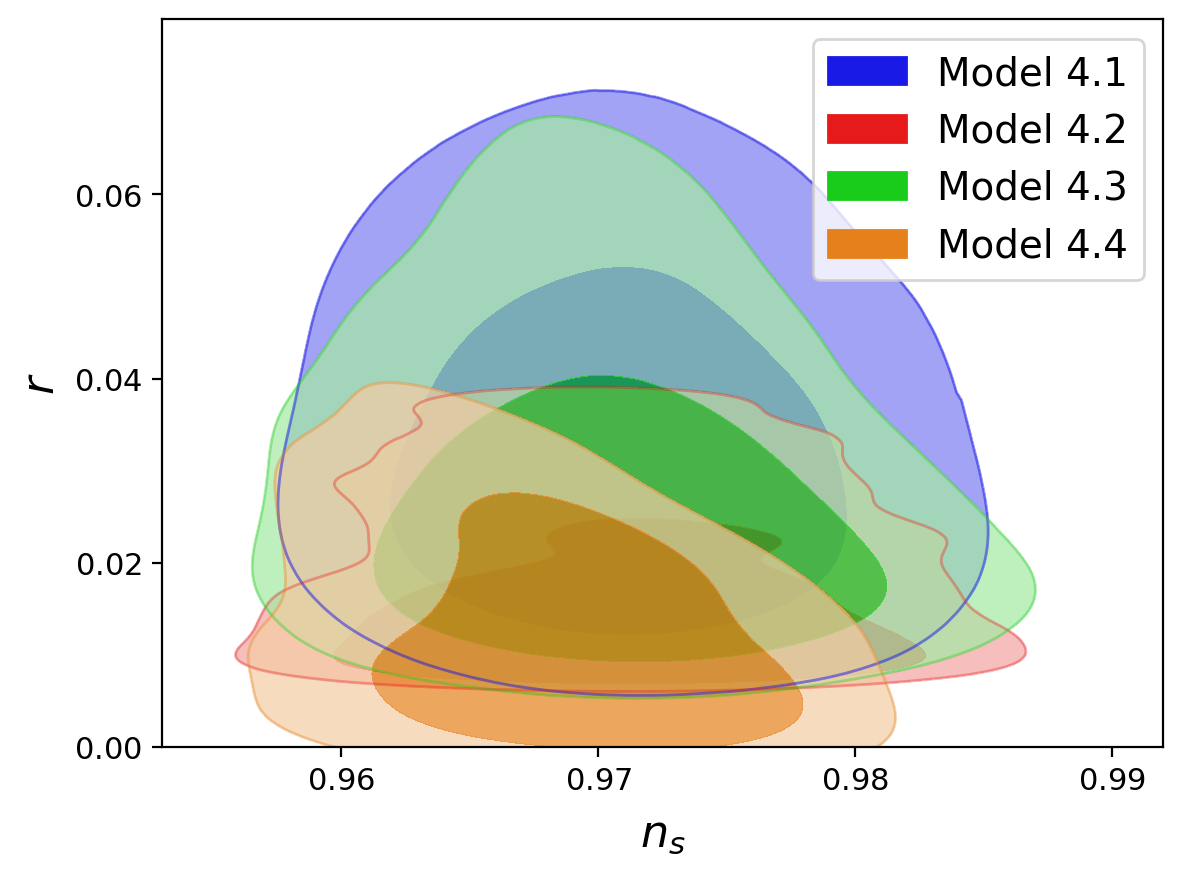} \\
        (c) & (d)
    \end{tabular}
    \caption{Marginalized constraints on the $(n_s,\, r)$ plane
    at $N = 60$ for all sixteen models grouped by Hubble
    background: (a)~de~Sitter $H = H_0$, models~1.1--1.4;
    (b)~quasi-de~Sitter $H = H_0 - H_1 t$, models~2.1--2.4;
    (c)~exponential $H = H_0\mathrm{e}^{-\Omega t}$,
    models~3.1--3.4; (d)~fractional $H = n/t$,
    models~4.1--4.4. Dark and light shaded regions correspond
    to $1\sigma$ (68.3\%) and $2\sigma$ (95.5\%) confidence
    levels respectively. Reference contours for Planck~2018
    and ACT~DR6.02 are shown in each panel.}
    \label{fig:allmodels}
\end{figure}

A systematic analysis of sixteen inflationary models within the
framework of ghost-free $f(R,\mathcal{G})$ revealed a number of
common patterns obtained as a result of Bayesian MCMC analysis
based on Planck and ACT data. A general summary of all sixteen
models combined by parameterizations of the Hubble parameter is
shown in Fig.~\ref{fig:allmodels}.

The first and most important result is that the preference for the
dataset is based not on the coupling function $h(\chi)$, but on
the choice of parametrization of the Hubble parameter. This fact
is especially noticeable when comparing the de Sitter background
$H = H_0$ and the flat quasi-de Sitter $H = H_0 - H_1 t$ case,
where the addition of $H_1 t$ shifted the values of the spectral
index of scalar perturbations $n_s$ to a bluer side and increased
the tensor-scalar ratio $r$. Such shifts have led to
configurations with the $H = H_0 - H_1 t$ parametrization
preferring the ACT DR6.02 data more. At the same time, exponential
parametrization of $H= H_0 e^{-\Omega t}$ leads to an equal
preference for both datasets. The fractional background $H = n/t$,
on the contrary, leads to a wide variation in preferences.

The second important result relates to the quality of the
selection of parameters. Exponential parameterization $H = H_0
e^{-\Omega t}$ demonstrates the best quality of the parameter
selection, as can be seen from the oval shapes of the contours
with large light areas. At the same time, all contours are grouped
in one place without a large spread of values. Therefore, such
parametrization is most organically combined with ghost-free
$f(R,\mathcal{G})$ by the model. On the contrary, the fractional
background $H = n/t$ is characterized by the least uniform
contours, wide confidence intervals of the parameter $n$ and the
presence of asymptotic lower bounds of the tensor-to-scalar ratio.

The third result is related to the role of the coupling function
$h(\chi)$. Despite the dominance of the parametrization of the
Hubble parameter in contour formation, the type of coupling
function has a significant impact on the tensor-to-scalar ratio
and on the quality of the selection of model parameters. It can be
seen that the power-law coupling demonstrates the most compact and
uniform contours for any parameterization. Exponential types of
coupling, especially with negative degrees, on the contrary,
demonstrate extremely rapid growth of the spectral index $n_s(N)$.
The hybrid coupling proposed in this paper eliminates this
disadvantage by reducing the growth rate and smoothing the
contours.

The fourth result is a steady reproduction of the values of the
parameter $\mu\approx 0.10$ , which is observed in each
configuration. It is important to note that this particular
parameter often has narrow confidence intervals, which implies its
better selection quality among other parameters. Finally, the
results of this work indicate the ability of ghost-free
$f(R,\mathcal{G})$ not only to support a wide range of
inflationary scenarios, but also to be able to match both
datasets. The analysis shows the prospects of the model. Future
research will focus on exploring aspects of the late expansion
using other observational data. Preliminary results obtained via
Noether symmetry analysis and dynamical systems methods indicate
the existence of stable solutions in the future Universe, with the
value $\mu \approx 0.10$ reproduced independently in the late-time
regime.

\begin{acknowledgments}
This research was funded by the Science Committee of the Ministry
of Science and Higher Education of the Republic of Kazakhstan
(Grant No. AP26194585).
\end{acknowledgments}

\end{document}